\documentclass[lettersize,journal]{IEEEtran}
\usepackage{amsmath,amsfonts}
\usepackage{mathtools}
\usepackage{steinmetz}
\usepackage{algorithmic}
\usepackage{algorithm}
\usepackage{array}
\usepackage[caption=false,font=normalsize,labelfont=sf,textfont=sf]{subfig}
\usepackage{makecell}

\usepackage{textcomp}
\usepackage{stfloats}
\usepackage{url}
\usepackage{verbatim}
\usepackage{graphicx}
\usepackage{cite}
\usepackage{xcolor}
\usepackage{hyperref}
\hypersetup{hidelinks=true}
\newtheorem{theorem}{Theorem}
\newtheorem{definition}{Definition}
\newtheorem{lemma}{Lemma}
\newtheorem{assumption}{Assumption}
\newtheorem{remark}{Remark}

\newtheorem{corollary}{Corollary}
\newcommand{\qed}{\hfill \ensuremath{\Box}}
\newcommand\norm[1]{\left\lVert#1\right\rVert}
\usepackage{tikz}
\usetikzlibrary{arrows, positioning, calc}
\usepackage{environ}
\newcommand*\squeezespaces[1]{
  \thickmuskip=\scalemuskip{\thickmuskip}{#1}
  \medmuskip=\scalemuskip{\medmuskip}{#1}
  \thinmuskip=\scalemuskip{\thinmuskip}{#1}
  \nulldelimiterspace=#1\nulldelimiterspace    
  \scriptspace=#1\scriptspace                  
}
\newcommand*\scalemuskip[2]{
  \muexpr #1*\numexpr\dimexpr#2pt\relax\relax/65536\relax
} 

\makeatletter
\newsavebox{\measure@tikzpicture}
\NewEnviron{scaletikzpicturetowidth}[1]{%
	\def\tikz@width{#1}%
	\def\tikzscale{1}\begin{lrbox}{\measure@tikzpicture}%
		\BODY
	\end{lrbox}%
	\pgfmathparse{#1/\wd\measure@tikzpicture}%
	\edef\tikzscale{\pgfmathresult}%
	\BODY
    }
\begin{document}

\title{Robust Performance Analysis and Nonlinearity Shaping for Closed-loop Reset Control Systems}

\author{S. Ali Hosseini, Dragan Kosti\'c, and S. Hassan HosseinNia,~\IEEEmembership{Senior Member~IEEE}
\thanks{This project is co-financed by ASMPT and Holland High Tech, Topsector High Tech Systems and Materials, with a PPS innovation grant public-private collaboration for research and development.}
\thanks{S. Ali Hosseini, and S. Hassan HosseinNia are with the Department of Precision and Microsystems Engineering, Delft University of Technology, 2628 CD Delft, The Netherlands (e-mail: S.A.Hosseini@tudelft.nl; S.H.HosseinNiaKani@tudelft.nl).}
\thanks{Dragan Kosti\'c is with ASMPT, 6641 TL Beuningen, The Netherlands (e-mail: dragan.kostic@asmpt.com).}}
\markboth{Journal of \LaTeX\ Class Files,~Vol.~14, No.~8, August~2021}%
{Shell \MakeLowercase{\textit{et al.}}: A Sample Article Using IEEEtran.cls for IEEE Journals}
\IEEEpubid{}
\maketitle
\begin{abstract}
Reset elements are nonlinear filters that improve control performance beyond linear time-invariant (LTI) limits but introduce higher-order harmonics that complicate design. Although frequency-domain tools like describing functions (DFs) and higher-order sinusoidal-input describing functions (HOSIDFs) analyze reset control systems (RCS), no direct method yet quantifies the impact of higher-order harmonics on the error signal without time-domain simulations. This paper introduces a robustness factor, $\sigma_2(\omega)$, which quantifies the increase in the root-mean-square (RMS) value of the error signal due to HOSIDFs, enabling RCS to rely solely on first-order DF characteristics while accounting for nonlinear effects. By using this robustness factor, a systematic method for designing pre- and post-filters is developed to ensure a predefined bound on $\sigma_2(\omega)$, thereby limiting the influence of higher-order harmonics without altering first-order DF behavior. The proposed framework is validated through a case study on a planar precision positioning stage, demonstrating how the robustness factor guides the reduction of nonlinearities and improves performance predictability.
\end{abstract}


\section{Introduction}\label{Sec: Introduction}
\IEEEPARstart{R}{ESET} elements are nonlinear filters used to overcome the fundamental performance limitations of linear time-invariant (LTI) control systems \cite{zhao2019overcoming}. The concept of reset control was initially introduced in \cite{clegg1958nonlinear} as a nonlinear integrator, later known as the Clegg integrator (CI).
In \cite{saikumar2019constant,guo2009frequency,hazeleger2016second}, reset control demonstrated promising behavior in overcoming the limitations inherent in linear feedback control caused by Bode’s gain-phase relationship \cite{freudenberg2000surveyBodeGain}.

In \cite{guo2009frequency}, the reset element is presented in the frequency domain using the describing function (DF) method. In \cite{saikumar2021loop}, the extension of the frequency domain tool ‘Higher Order Sinusoidal-Input Describing Functions (HOSIDFs)’ for reset controllers is introduced, enabling deeper open-loop analysis. Additionally, in \cite{saikumar2021loop,LukeLure}, a method for translating open-loop behavior to closed-loop in the frequency domain is proposed, which corresponds to the HOSIDF-based sensitivity functions for reset control systems (RCS). It provides an indication of how reliably the DF-based reset control design predicts system behavior.


Several studies have employed the HOSIDFs tool to shape higher-order harmonics while maintaining the DF either unchanged or within a desirable range \cite{karbasizadeh2022band, karbasizadeh2022continuous, LukeIFAC2024}.

In \cite{cai2020optimal}, it is shown that by carefully selecting the sequence of loop components, one can reduce the effect of HOSIDFs without affecting the DF. In \cite{karbasizadeh2022continuous}, a pre- and post-filtering method was proposed, whereby the first-order harmonic of the reset control system remains unchanged while allowing the shaping of HOSIDFs.
Furthermore, a comparable approach was adopted in \cite{ZHANG2024106063}, this time in the closed-loop, where the authors utilized the HOSIDF-based sensitivity function to minimize the ratio of the third-order sensitivity to the first-order sensitivity, thereby reducing the influence of higher-order harmonics.

Despite these advancements, there remains a gap in the development of a higher-order harmonics quantifier capable of directly revealing the effects of all HOSIDFs on the error signal without requiring time-domain analysis. Such an observer would enhance our understanding of the reliability of DF-based analyses. Therefore, as the first contribution of this study:

\begin{itemize}
\item A robustness factor is introduced that integrates higher-order harmonics with the fundamental harmonic using the 2-norm. This factor provides a metric for estimating the additional error introduced by the reset element’s nonlinearity during the design process. Importantly, it can be computed entirely in the frequency domain using closed-loop HOSIDFs, thereby significantly reducing the computational cost of predicting the time-domain behavior of reset control systems.
\end{itemize}




In all the above-mentioned methods \cite{karbasizadeh2022band, karbasizadeh2022continuous, LukeIFAC2024, cai2020optimal}, the reduction of the HOSIDFs was achieved by modifying reset control systems with predefined pre- and post-filter structures. However, the reduction obtained with these filters was often insufficient, requiring multiple iterations to identify a suitable filter and consequently achieve the desired performance. Thanks to the first contribution of this study, this process is now streamlined. As the second contribution of this study:

\begin{itemize}
    \item A method is proposed for designing pre- and post-filters, without imposing a fixed filter structure, by constraining the robustness factor below a specified threshold. This ensures that the nonlinearity of the error signal remains within acceptable limits while preserving the DF characteristics and explicitly reducing the HOSIDFs.
\end{itemize}

Thus, by employing the methods introduced in this study (the two aforementioned contributions), it is possible to first predict the influence of higher-order harmonics on the error signal using the robustness factor. Given a desired robustness level, the required pre- and post-filters can then be computed to shape the system's nonlinearities such that the robustness factor remains within a specified bound.

The remainder of this paper is structured as follows. In Section \ref{sec: Preliminaries}, preliminaries on reset control in the time and frequency domains are given. In Section \ref{sec: case study}, an illustrative example is presented to highlight existing challenges and to demonstrate the impact of higher-order harmonics on the performance of reset control systems. Section \ref{sec: Robustness factor} introduces the proposed robustness factor, which serves to quantify the effect of HOSIDFs on the error signal. Section \ref{sec: shaping nonlinearity} presents a method to exploit the robustness factor for the robust design of reset control systems. This section also includes experimental results that demonstrate the effectiveness of the proposed filtering approach in reducing higher-order harmonics. Finally, conclusions and suggestions for future work are presented in Section \ref{sec: Conclusion}.


\section{Preliminaries}\label{sec: Preliminaries}
\subsection{Reset Control System}\label{SubSec: Reset element}  
Consider the closed-loop reset control system, illustrated in Fig. \ref{Fig: Block diagram CL}. In this setup, the exogenous inputs are given by \( (r(t), d_i(t), d_n(t)) \in \mathbb{R} \), while the control signal is represented as \( u(t) \in \mathbb{R} \), and the system’s output as \( y(t) \in \mathbb{R} \), all defined for \( t \in \mathbb{R}_{\geq 0} \). The plant, denoted as \( G \), operates under the influence of LTI filters, \( C_\text{pre} \), \( C_\text{par} \) and \( C_\text{pos} \). The system also incorporates a reset element, denoted by \( \mathcal{R} \), which is characterized as follows:
\begin{equation}
\label{eq: reset state space}
\mathcal{R} : 
\begin{cases}
    \dot{x}_r(t) = A_r x_r(t) + B_r e_r(t), & \text{if } \left(x_r(t), e_r(t)\right) \notin \mathcal{F}, \\[5pt]
    x_r(t^+) = A_\rho x_r(t), & \text{if } \left(x_r(t), e_r(t)\right) \in \mathcal{F}, \\[5pt]
    u_r(t) = C_r x_r(t) + D_r e_r(t), &
\end{cases}
\end{equation}
where the reset condition is governed by the reset surface \( \mathcal{F} \), defined as:
\begin{equation}
\label{reset surface}
\mathcal{F} := \{ e_r(t) = 0 \wedge (A_\rho - I) x_r(t) \neq 0 \}.
\end{equation}
Here, \( x_r(t) \in \mathbb{R}^{n_r\times 1} \) represents the state vector before reset, while \( x_r(t^{+})\in\mathbb{R}^{n_r\times 1} \) corresponds to the state immediately after reset. The reset element's state-space representation consists of matrices \( A_r \in \mathbb{R}^{n_r \times n_r} \), \( B_r \in \mathbb{R}^{n_r \times 1} \), \( C_r \in \mathbb{R}^{1 \times n_r} \), and \( D_r \in \mathbb{R} \). The reset matrix, responsible for modifying the state upon reset, is given by \( A_\rho = \text{diag}(\gamma_1, \dots, \gamma_{n_r}) \), where each \( \gamma_{i} \in (-1,1) \) for all \( i \in \mathbb{N} \). The reset element receives an input \( e_r(t) \in \mathbb{R} \) and produces an output \( u_r(t) \in \mathbb{R} \). 

When no reset event occurs, i.e., when \( (x_r(t), e_r(t)) \notin \mathcal{F} \) for all \( t \in \mathbb{R}_{\geq 0} \), the reset element follows the dynamics of its base linear system (BLS). The transfer function corresponding to this BLS is expressed as:
\begin{equation}
\label{eq RCS bls}
R_\text{bl}(s) = C_r(sI - A_r)^{-1}B_r + D_r,
\end{equation}
where \( s \in\mathbb{C} \) represents the Laplace variable.

\usetikzlibrary {arrows.meta}
\tikzstyle{block} = [draw,thick, fill=white, rectangle, minimum height=2em, minimum width=2.5em, anchor=center]
\tikzstyle{sum} = [draw, fill=white, circle, minimum height=0.6em, minimum width=0.6em, anchor=center, inner sep=0pt]

\usetikzlibrary {arrows.meta}
\tikzstyle{block} = [draw,thick, fill=white, rectangle, minimum height=2em, minimum width=2.5em, anchor=center]
\tikzstyle{sum} = [draw, fill=white, circle, minimum height=0.6em, minimum width=0.6em, anchor=center, inner sep=0pt]
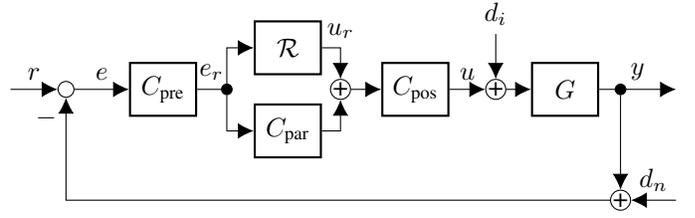
\begin{figure}[!t]
	\centering
	\begin{scaletikzpicturetowidth}{\linewidth}
		\begin{tikzpicture}[scale=\tikzscale]
			\node[coordinate](input) at (0,0) {};
			\node[sum] (sum1) at (1,0) {};
			\node[sum] (sum3) at (8.75,0) {+};
			\node[sum] (sum4) at (11,-2) {+};
			\node[sum, fill=black, minimum size=0.4em] (dot2) at (11,0) {};		
			\node[sum] (sumN) at (5.95,0) {+};
			\node[block] (lead) at (5.0,0.75) {$\mathcal{R}$};

\node[block] (C_par) at (5.0,-0.75) {$C_\text{par}$};
            
			\node[block] (controller) at (7.3,0) {$C_\text{pos}$};
			\node[block] (fo-higs) at (2.75,0) {$C_\text{pre}$};
			\node[block] (system) at (10,0) {$G$};

            \node[sum, fill=black, minimum size=0.4em] (Dot_1) at (3.9,0) {};
			\node[coordinate](output) at (12,0) {};
			\node[coordinate](di-input) at (8.75,1) {};
			\node[coordinate](n-input) at (12,-2) {};
			\draw[arrows = {-Latex[width=6pt, length=6pt]}] (input)  -- node[above]{$r$} (sum1);
			\draw[arrows = {-Latex[width=6pt, length=6pt]}] (di-input)node[above]{$d_i$}  --  (sum3);
			\draw[arrows = {-Latex[width=6pt, length=6pt]}] (n-input)  -- node[above]{$d_n$} (sum4);
			\draw[arrows = {-Latex[width=6pt, length=6pt]}] (sum1)   --node[above]{$e$}  (fo-higs);
			\draw[ = {-Latex[width=6pt, length=6pt]}] (fo-higs) --node [above]{$\,e_r$} (Dot_1);
			\draw[arrows = {-Latex[width=6pt, length=6pt]}] (lead)  -|node[above] {$u_r$} (sumN);

            \draw[arrows = {-Latex[width=6pt, length=6pt]}] (C_par)  -|node[above] {} (sumN);
            
			\draw[arrows = {-Latex[width=6pt, length=6pt]}] (controller)  --node[above] {$u$} (sum3);
			\draw[arrows = {-Latex[width=6pt, length=6pt]}] (sum3) -- (system);
			\draw[arrows = {-Latex[width=6pt, length=6pt]}] (system)  -- node[above]{$y$} (output);
			\draw[arrows = {-Latex[width=6pt, length=6pt]}] (dot2) -- (sum4);

\draw[arrows = {-Latex[width=6pt, length=6pt]}] (sumN) -- (controller);
            
			\draw[arrows = {-Latex[width=6pt, length=6pt]}] (sum4) -| node[pos=0.9,left]{$-$} (sum1);

            \draw[arrows = {-Latex[width=6pt, length=6pt]}] (Dot_1) |-node[pos=0.85,left]{} (lead);

            \draw[arrows = {-Latex[width=6pt, length=6pt]}] (Dot_1) |-node[pos=0.85,left]{} (C_par);
			
		\end{tikzpicture}
	\end{scaletikzpicturetowidth}
	\caption{Block diagram of the closed-loop reset control system.}
	\label{Fig: Block diagram CL}	
\end{figure}



\subsection{Frequency Domain Analysis of Reset Control Systems}\label{subsec: frequency reset}  
The nonlinear nature of reset elements poses significant challenges in controller design within the frequency domain, particularly when utilizing the widely adopted loop-shaping technique, which depends on Bode plots. Among the limited methods available for analyzing nonlinear controllers in the frequency domain, the describing function approach is one of the most prominent. This method approximates the steady-state behavior of a convergent nonlinear system by capturing its first harmonic component through a Fourier series expansion \cite{khalil2002nonlinear}.  

In the following, we introduce the open-loop and closed-loop frequency domain responses of reset control systems using the describing function method.

\subsubsection{Open-Loop Frequency Domain Analysis of Reset Control Systems}\label{subsec: openloop frequency}  

The describing function method has been employed in \cite{guo2009frequency} to represent reset elements in the frequency domain. Expanding on this approach, \cite{saikumar2021loop} introduced an extension of this frequency-domain tool known as HOSIDFs for reset controllers, enabling a more detailed open-loop analysis.  

Considering an open-loop system where the input is given by \( e(t,\omega) = \hat{e} \sin(\omega t) \), the corresponding output \( y(t,\omega) \) (with the loop remaining open) can be expressed as a Fourier series:  
\begin{equation}
    \label{eq: e to y}
y(t,\omega)=\sum_{n=1}^{\infty}\left|\mathcal{L}_n(j\omega)\right|\hat{e}\sin\left(n\omega t +\angle \mathcal{L}_n(j\omega)\right),
\end{equation}
where \( n \in \mathbb{N} \), and \( \mathcal{L}_n(j\omega) \) $\forall n\neq1$ is given by (see \cite[equation (34)]{LukeLure}):  
\begin{equation}
    \label{eq: Ln open loop}
\mathcal{L}_{n}(\omega)=G(nj\omega)C_\text{pos}(nj\omega)H_n(\omega)C_\text{pre}(j\omega)e^{j(n-1)\angle C_\text{pre}(j\omega)},
\end{equation}
and for $n=1$ we have (\cite[equation (30)]{LukeLure})
\begin{equation}
    \label{eq: L1 open loop}
\mathcal{L}_{1}(\omega)=G(j\omega)C_\text{pos}(j\omega)[H_1(\omega)+C_\text{par}(j\omega)]C_\text{pre}(j\omega).
\end{equation}

The \( H_n(\omega) \) represents the HOSIDF of the reset element \( \mathcal{R} \), introduced in \cite[Theorem 3.1]{saikumar2021loop}.

Although the expressions in \eqref{eq: Ln open loop} and \eqref{eq: L1 open loop} represent the first-order and higher-order DF-based open-loop frequency responses of the reset control system, the presence of a nonlinear component within the loop prevents closed-loop frequency domain functions—such as sensitivity and complementary sensitivity—from adhering to the standard open-loop/closed-loop relationships observed in LTI systems. Consequently, direct computation of the closed-loop HOSIDFs is essential for analyzing reset control systems.\\

\subsubsection{Closed-loop Frequency Domain Analysis of Reset Control Systems}
As previously discussed, unlike LTI systems, there is no direct correspondence between the open-loop and closed-loop frequency domain responses for nonlinear controllers, particularly in the case of reset control systems. Therefore, it is desirable to predict closed-loop performance using only the frequency domain characteristics of the loop components, including the plant. \cite{LukeLure} provides a frequency domain-based approach for performance prediction in RCSs under the following assumptions.
\begin{assumption}
    \label{Assumption reset convergence}
    \cite[Assumption 2.1]{LukeLure}, For reset element $\mathcal{R}$ as in \eqref{eq: reset state space}, it holds that
    \begin{equation}
\left|\lambda\left(A_\rho e^{A_r\delta}\right)\right|<1 \quad \forall\delta\in \mathbb{R}_{>0},
    \end{equation}
    where $\lambda(\cdot)$ represents the eigenvalues of a matrix.
\end{assumption}

It is shown in \cite[Proposition 2]{guo2009frequency}, the output of
the reset element subject to the sinusoidal
input $e_r(t)=\hat{e}_r\sin{\left(\omega t+\varphi_{e_r}\right)}$ (with $\hat{e}_r\in \mathbb{R}_{>0}$ and $\varphi_{e_r} \in \mathbb{R}$) converges to a $\frac{2\pi}{\omega}$ periodic solution when
Assumption \ref{Assumption reset convergence} holds. Note that the condition in Assumption \ref{Assumption reset convergence} is
satisfied if the base-linear system is stable ($A_r$ is Hurwitz)
and reset matrix $A_\rho$ is a Schur stable matrix ($|\lambda(A_\rho)| < 1$)
\cite[Remark 4]{guo2009frequency}. However, it is also possible to satisfy the assumption if these conditions are not satisfied.\\

\begin{assumption}
    \label{Assumption: closed-loop convergence}
    \cite[Assumption 3.1]{LukeLure}, Consider the system in Fig. \ref{Fig: Block diagram CL} with any
initial conditions for the LTI part of the closed-loop system and reset element $\mathcal{R}$. Then, given a sinusoidal external input as $w(t)=\hat{w}\sin{\left(\omega t+\varphi_\omega\right)}$ (with $\hat{w}\in \mathbb{R}_{>0}$ and $\varphi_\omega \in \mathbb{R}$), the signals $u_r$, $e$ and $e_r$ converge to a unique periodic solution with period $T = \frac{2\pi}{\omega}$ and a zero-mean (as in \cite[Definition 3.1]{LukeLure}).
\end{assumption}
This assumption ensures that a unique periodic solution exists for the output of the reset element, which shares the same period as the reset input. Consequently, the reset output signal can be represented using a Fourier series. The validity of this assumption can be evaluated using the FRF-based method presented in \cite[Corollary 1 and Theorem 2]{hosseini2025frequency}.

Having satisfied Assumption \ref{Assumption reset convergence} and \ref{Assumption: closed-loop convergence}, by considering $r(t)=\hat{r}\sin{\left(\omega t\right)}$, $d_i=0$, and $d_n=0$, for error signal $e(t,\omega)$ we can write:
\begin{equation}
    \label{eq ess}
        e_\text{}(t,\omega)=\sum_{n=1}^{\infty}e_n(t,\omega),
\end{equation}
    where
\begin{equation}
    \label{eq en}
e_n(t,\omega)=|S_{r,e}^{n}(\omega)|\sin{(n\omega t+\angle{S_{r,e}^{n}(\omega)})},
\end{equation}
with higher-order sinusoidal input sensitivity functions $S_{r,e}^{n}(\omega)$ ($\text{n}^\text{th}$-order HOSIDF calculated from input $r$ to the output e) as \cite[equations (32)-(34)]{LukeLure}:
\begin{equation}
\label{eq Sn}
S_{r,e}^{n}(\omega) =
\begin{cases}
\displaystyle \frac{1}{1 + \mathcal{L}_1(\omega)}, \hfill \text{for } n = 1, \\[10pt]
\displaystyle -\mathcal{L}_n(\omega) S_\mathrm{bl}(nj\omega) 
\left(|S_{r,e}^{1}(\omega)| e^{jn\angle S_{r,e}^{1}(\omega)}\right), \\
\hfill \text{for odd } n \geq 2, \\[5pt]
0, \hfill \text{for even } n \geq 2,
\end{cases}
\end{equation}
 where $S_\mathrm{bl}(nj\omega)=\frac{1}{1+L_\mathrm{bl}(nj\omega)}$ where $L_\mathrm{bl}(j\omega)=G(j\omega)C_\text{pos}(j\omega)[C_\text{par}(j\omega)+R_\text{bl}(j\omega)]C_\text{pre}(j\omega)$ is the base linear transfer function of the open-loop. For simplicity, we will write \( e(t) \) instead of \( e(t, \omega) \), and apply the same convention to other sinusoidal functions.\\

The existence of closed-loop HOSIDFs enables a more precise design of reset control systems. However, analyzing reset controllers while accounting for all harmonic components is a non-trivial task, due to the presence of infinitely many higher-order sensitivities. This complexity poses significant challenges in drawing clear conclusions about their impact on system performance. In the following section, we introduce a closed-loop reset control system as a case study, allowing us to analyze the influence of both open-loop and closed-loop HOSIDFs on the error signal.

\section{Case study and problem definition: a planar precision positioning stage control}\label{sec: case study}
\begin{figure}[!t]
\centering
\includegraphics[width=0.6\columnwidth]{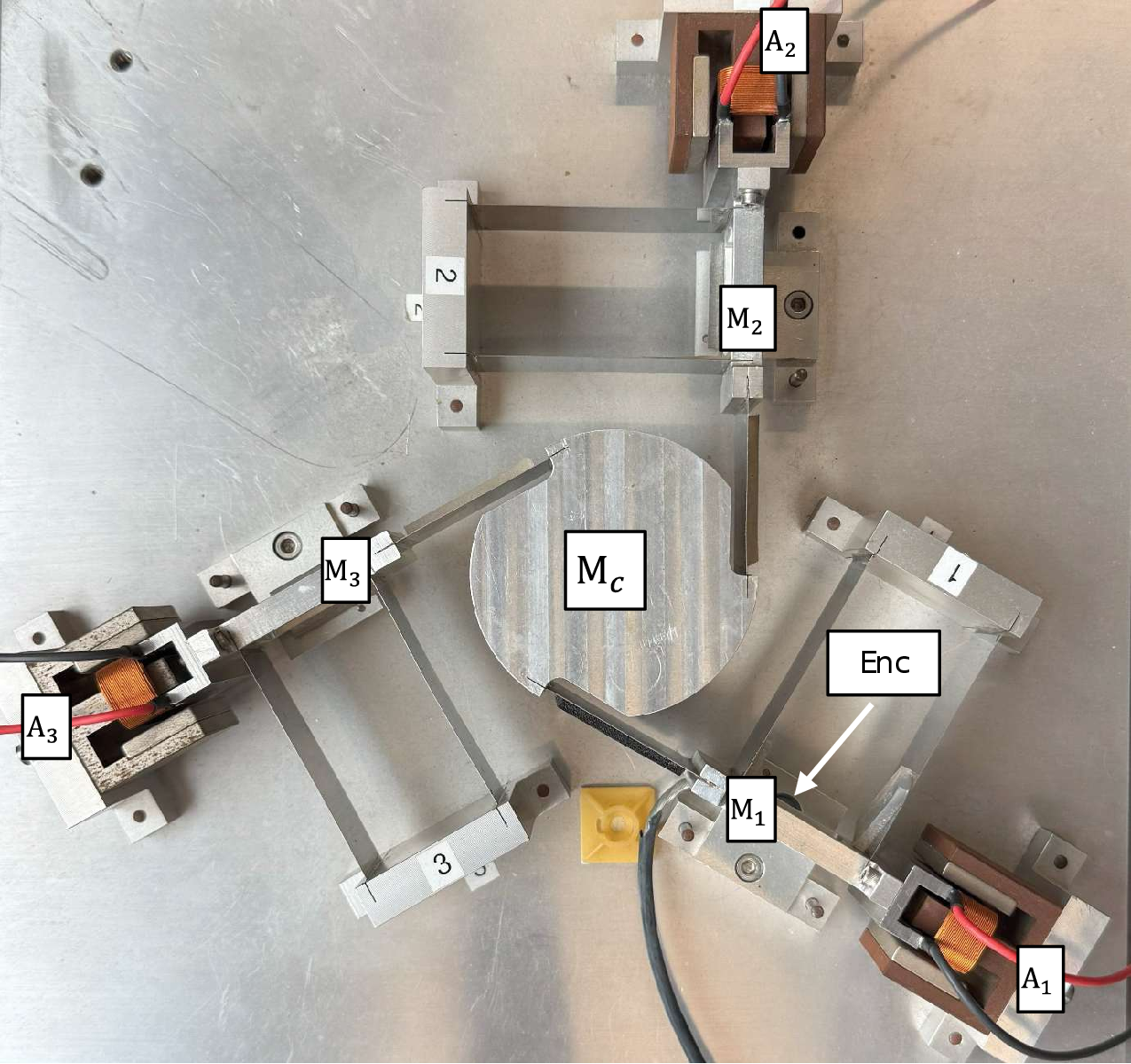}
\caption{Planar precision positioning ‘Spider’ stage with voice coil actuators, denoted as $\text{A}_1$, $\text{A}_2$, and $\text{A}_3$, that control three moving masses, labeled $\text{M}_1$, $\text{M}_2$, and $\text{M}_3$. These masses are constrained by leaf flexures. The central mass, denoted as $\text{M}_\text{c}$, is connected to $\text{M}_1$, $\text{M}_2$, and $\text{M}_3$ via additional leaf flexures. Position feedback is provided by linear encoders (Mercury M2000), labeled as Enc, which are placed beneath masses $\text{M}_1$, $\text{M}_2$, and $\text{M}_3$ \cite{saikumar2021loop}.}
\label{Fig: Spider physical}
\end{figure}

\begin{figure}[!t]
\centering
\includegraphics[width=0.95\columnwidth]{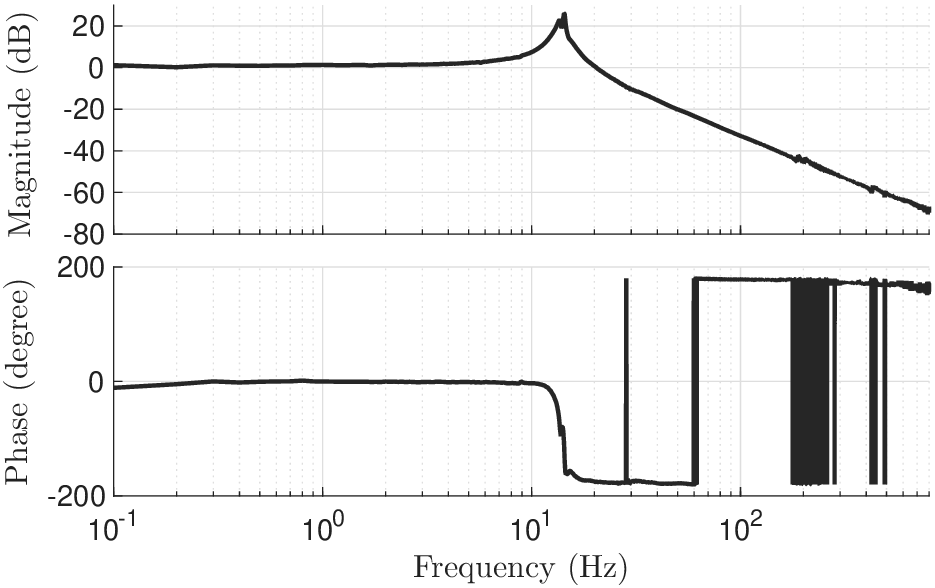}
\caption{The FRF data of the planar precision positioning stage, illustrating the
mapping of actuator force from $\text{A}_1$ to the displacement measured by the encoder in $\text{M}_1$ location.}
\label{Fig: FRF spider}
\end{figure}

Considering the closed-loop system shown in Fig.~\ref{Fig: Block diagram CL}, the planar motion system with three degrees of freedom, illustrated in Fig.~\ref{Fig: Spider physical}, is taken as the physical system under study. For the purpose of SISO reset control design, the plant \( G \) is defined as the transfer function from the force applied by actuator \( \text{A}_1 \) to the resulting displacement of mass \( \text{M}_1 \). The corresponding measured frequency response function is shown in Fig. \ref{Fig: FRF spider}.

\subsubsection{Linear and Reset Controller Design}
We consider a PID controller (including a low-pass filter) of the form  
\begin{equation}
	\label{PID1}
	C_{\text{PID}}(s) = k_p\left(1 + \frac{\omega_{i}}{s}\right)\left(\frac{1 + \frac{s}{\omega_d}}{1 + \frac{s}{\omega_t}}\right)\left(\frac{1}{1+\frac{s}{\omega_\text{LF}}}\right),
\end{equation}
where \(k_p \in \mathbb{R}\) denotes the proportional gain, which is tuned to ensure a 0 dB gain at the bandwidth frequency \(\omega_c \in \mathbb{R}_{>0}\) (i.e., the crossover frequency). The parameter \(\omega_i \in \mathbb{R}_{>0}\) specifies the frequency at which integral action is stopped, \(\omega_d \in \mathbb{R}_{>0}\) marks the onset of the derivative action, \(\omega_t \in \mathbb{R}_{>0}\) determines the frequency at which the derivative action is attenuated, and $\omega_\text{LF}\, \in \mathbb{R}_{>0}$ is the corner frequency of the low-pass filter used to suppress high-frequency noise and other unwanted components. Assuming a purely linear controller, i.e., \(C_{\text{L}} = C_{\text{PID}}\), the PID parameters are selected to achieve a bandwidth of \(100 \times 2\pi\) rad/s and a phase margin of \(30^\circ\), as summarized in Table \ref{tab: controller parameters}.

\begin{table}[]
\caption{Parameters for $C_\text{L}$ and $C_\text{NL}$ (all frequencies are presented in Hz).}
\label{tab: controller parameters}
\resizebox{1\columnwidth}{!}{%
\begin{tabular}{ccccccccc}

          & \large{$k_p$} &\large{$\omega_i$} & \large{$\omega_d$} & \large{$\omega_t$}& \large{$\omega_\mathrm{LF}$}& \large{$\omega_l$} &\large{$\omega_f$}           &\large{$A_\rho$}    \\ \hline
\large{$C_{\mathrm{L}}$}       & \large{$14.79$}        &\large{$17.86$}& \large{$33.33$}                &\large{$300$}&\large{$423$}& $-$ &$-$         & $- $ \\ 
\large{$C_{\mathrm{NL}}$}       & \large{$13.62$}        &\large{$46.51$}& \large{$33.33$}                &\large{$300$}&\large{$423$}&\large{$80$} &\large{$350$}         & \large{$0$} \\ \hline
\end{tabular}
}
\end{table}

As established in control theory, improving tracking performance and disturbance rejection requires increasing the controller gain at low frequencies. However, according to Bode's gain–phase relationship \cite{freudenberg2000surveyBodeGain} and the waterbed effect \cite{waterbed2012}, this increase in low-frequency gain typically results in a reduction of phase margin at the bandwidth frequency. 

To address this issue, we employ reset control to compensate for the phase loss induced by the gain increase at low frequencies. Specifically, we consider a Constant-in-Gain, Lead-in-Phase (CgLp) element designed to provide an additional \(15^\circ\) of phase at the bandwidth frequency. This allows us to increase the integral frequency \(\omega_i\) without compromising the phase margin, thereby enhancing the low-frequency gain. In this study, we adopt the CgLp structure proposed in \cite{hosseini2025AddOnFilterDesign}, defined as follows:

\usetikzlibrary {arrows.meta}
\tikzstyle{sum} = [draw, fill=white, circle, minimum height=0.0em, minimum width=0.0em, anchor=center, inner sep=0pt]
\tikzstyle{block} = [draw,thick, fill=white, rectangle, minimum height=2.5em, minimum width=3em, anchor=center]
\begin{figure}[!t]
	\centering
	\begin{scaletikzpicturetowidth}{0.97\linewidth}
		\begin{tikzpicture}[scale=\tikzscale]
			\node[coordinate](input) at (0,0) {};
			
			\node[sum] (sum3) at (8.5,0) {};
			

			\node[block] (lead) at (3.2,0) {${k_\text{c}}$};
			\node[block] (controller) at (4.9,0) {$C_\mathfrak{c}$};
			\node[block] (fo-higs) at (1.5,0) {$\mathcal{R}$};

            \node[block] (PID) at (7.0,0) {$C_\text{PID}$};

			\node[coordinate](output) at (9.5,0) {};

			\draw[arrows = {-Latex[width=8pt, length=10pt]}] (input)  -- node[above]{} (fo-higs);
			\draw[arrows = {-Latex[width=8pt, length=10pt]}] (fo-higs) --node [above]{} (lead);
			\draw[arrows = {-Latex[width=8pt, length=10pt]}] (lead)  --node[above] {} (controller);
			\draw[arrows = {-Latex[width=8pt, length=10pt]}] (controller)  --node[above] {} (PID);

            \draw[arrows = {-Latex[width=8pt, length=10pt]}] (PID)  --node[above] {} (sum3);

              \draw [color=gray,thick,dotted](0.4,-0.75) rectangle (5.7,0.75);
			\node at (0.3,1) [right,color=gray] {$\text{CgLp}$};

            \draw [color=gray,thick,dashed](0.2,-1.2) rectangle (7.8,1.4);
			\node at (0.3,1.55) [left,color=gray] {$C_\text{NL}$};
			
		\end{tikzpicture}
	\end{scaletikzpicturetowidth}
	\caption{The reset-based controller structure (including CgLp).}
	\label{Fig: Block diagram CgLp2}	
\end{figure}
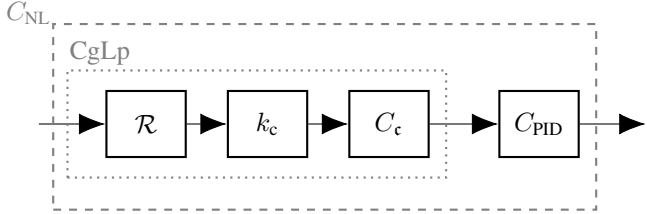

\begin{definition}
    \label{Definition: New CgLp}
    \cite[Definition 1]{hosseini2025AddOnFilterDesign}, We construct the CgLp filter as illustrated in Fig. \ref{Fig: Block diagram CgLp2}, where
     \begin{equation}\label{eq: kc}
        k_\text{c} = \frac{\omega_f - \omega_l}{\omega_f},
    \end{equation}
    \begin{equation}\label{eq: Cc}
        C_\mathfrak{c}(s) = \frac{1 + s / \omega_l}{1 + s / \omega_f},
    \end{equation}
    with \( [\omega_l, \omega_f] \in \mathbb{R}^{1\times 2}_{>0} \), and \( \mathcal{R} \) is a proportional GFORE element ($n_r=1$) characterized by
    \begin{equation} \label{eq: Dr}
         A_r = -\omega_r, \quad B_r = 1, \quad C_r = \omega_r, \quad D_r = \frac{\omega_l}{\omega_f - \omega_l},
    \end{equation}
     with $\omega_r\in \mathbb{R}_{>0}$ as
    \begin{equation}
    \label{eq: corner frequency}
    \omega_r = \frac{\omega_l}{\sqrt{1 + \left(\frac{4(1 - A_\rho)}{\pi(1 + A_\rho)}\right)^2}}.\\
    \end{equation}
\end{definition}

According to Definition \ref{Definition: New CgLp} and \cite[Theorem 3]{hosseini2025AddOnFilterDesign}, the parameters of the CgLp element required to achieve an additional \(15^\circ\) phase lead at the bandwidth frequency are calculated and summarized in Table \ref{tab: controller parameters}. By incorporating the CgLp element into the control loop and defining the nonlinear controller as \(C_{\text{NL}} = C_{\text{PID}} \times \text{CgLp}\) (see Fig. \ref{Fig: Block diagram CgLp2}), we are able to increase the integral frequency from \(\omega_i = 17.8\times 2\pi\) to \(\omega_i = 46.5\times 2\pi\) (more than 2.5 times higher). This adjustment enhances low-frequency gain while maintaining the desired phase margin. Please note that for $C_\text{NL}$ it is considered $C_\text{pre}=1$, $C_\text{par}=0$, and $C_\text{pos}=k_\text{c}C_\text{PID}C_\mathfrak{c}$.

\begin{figure}[!t]
\centering
\includegraphics[width=0.9\columnwidth]{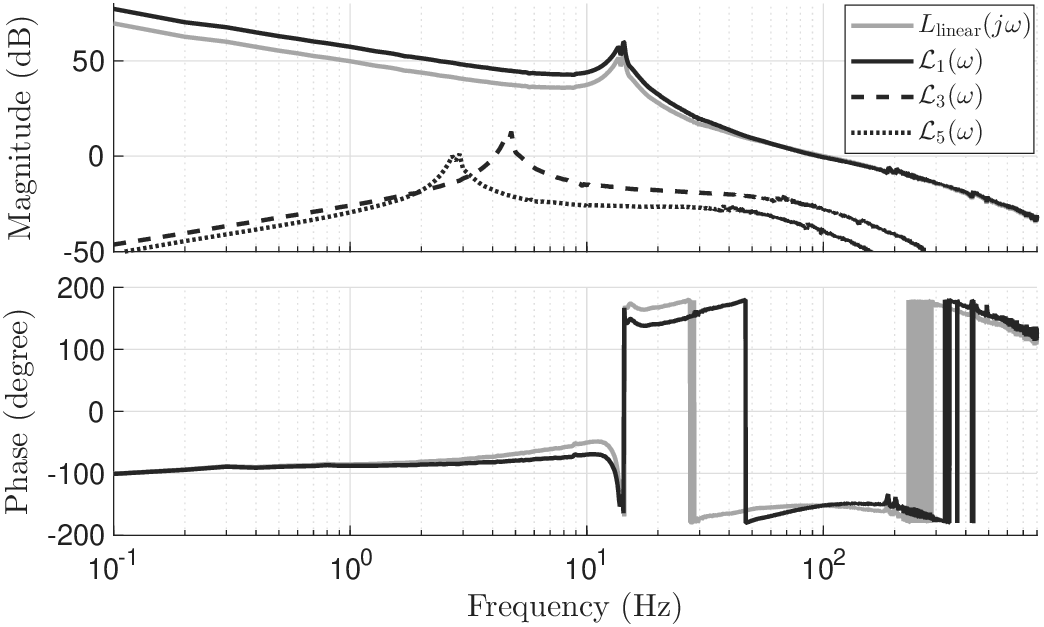}
\caption{Open-loop frequency response of the linear controller with the first-, third-, and fifth-order open-loop DF of the nonlinear controller.}
\label{Fig: Ln}
\end{figure}

In Fig. \ref{Fig: Ln}, the open-loop transfer function of the linear controller ($L_\text{linear}(s)=G(s)C_\text{L}(s)$) is presented alongside the first ($\mathcal{L}_{1}(\omega)$, as in \eqref{eq: L1 open loop}) and higher-order DFs ($\mathcal{L}_{n}(\omega)$, as in \eqref{eq: Ln open loop}) of the open-loop frequency response for the nonlinear controller. It can be observed that the reset-based controller provides higher gain at low frequencies without affecting the system's bandwidth or phase margin. This characteristic facilitates improved tracking and disturbance rejection.

\begin{figure}[!t]
\centering
\includegraphics[width=0.9\columnwidth]{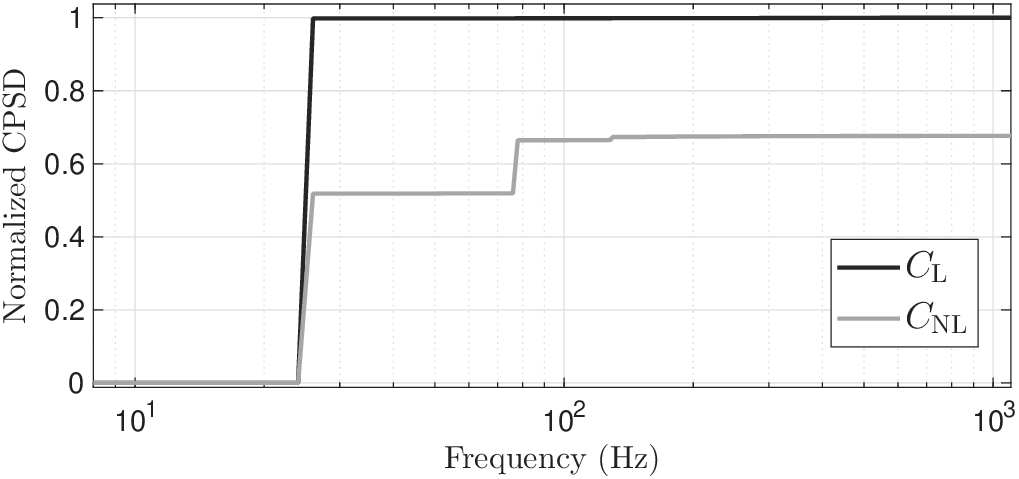}
\caption{The PSD of the measured error signal for both linear and nonlinear control.}
\label{Fig: PSD PID and reset}
\end{figure}

To validate this, both linear controller ($C_\text{L}$) and reset-based controller ($C_\text{NL}$) are implemented on the physical setup (Fig. \ref{Fig: Spider physical}) for a reference tracking case where the reference is considered as a sinusoidal input at frequency $f_r$ as $r(t)=\hat{{r}}\sin{(2\pi f_r t)}$. The cumulative power spectrum density (CPSD) of the errors in both linear and nonlinear cases is plotted in Fig. \ref{Fig: PSD PID and reset}, for $f_r=28\,$Hz and $\hat{{r}}=32\,\mu \text{m}$ (400 steps of encoder with resolution of $0.08$ $\mu$m). It can be observed that, as expected, the reset-based controller outperforms the linear controller due to its higher gain at low frequencies. However, the CPSD of the $C_\text{NL}$ exhibits noticeable jumps, which contribute to higher error and, consequently, an increased root mean square (RMS) error. These jumps are attributed to higher-order harmonics introduced by the reset element, as they occur at odd multiples (three, five, etc.) of the fundamental frequency. Although there is a clear reduction in tracking error compared to the linear case, to further optimize the nonlinear controller, it is important to predict the impact of these higher-order harmonics on the error signal prior to implementation. This would enable a redesign of the reset controller to minimize the influence of these harmonics on the error signal.

\begin{figure}[!t]
\centering
\includegraphics[width=0.95\columnwidth]{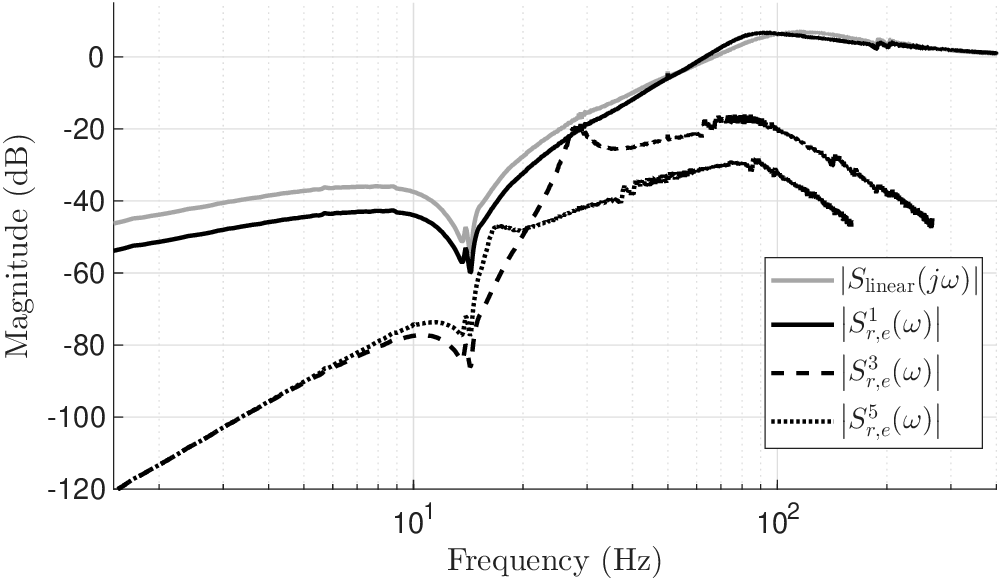}
\caption{Closed-loop sensitivity frequency response of the linear controller with the first-, third-, and fifth-order sensitivity of the nonlinear controller.}
\label{Fig: Sn}
\end{figure}

Therefore, to provide a clearer indication of the closed-loop system behavior, the first- and higher-order DF-based sensitivities of the nonlinear controller, as defined in \eqref{eq Sn}, are shown in Fig. \ref{Fig: Sn} alongside the sensitivity function of the linear controller. As we can see,  there are higher-order harmonics existing in both open-loop and closed-loop, and relying on only first-order DF ($\mathcal{L}_{1}(\omega)$ and $S_{r,e}^{1}(\omega)$) might result in unpredictable errors. Although these HOSIDFs offer valuable insights into the presence and effects of nonlinearity in RCSs, it remains unclear to what extent, and in what manner, these higher-order components should be shaped or attenuated to justify relying solely on the first-order harmonic.



Thus, many studies aim to ensure that the HOSIDFs remain well separated from the first-order DF, as seen in~\cite[Section 4]{LukeIFAC2024}, \cite{karbasizadeh2022band}, and \cite[Section III]{hosseini2025AddOnFilterDesign}, where the reset control system is designed to position the open-loop third-order harmonic far from the first-order harmonic. In~\cite{zhang2024enhancing}, efforts are made to minimize the ratio $\frac{\left|S_{r,e}^{3}(\omega)\right|}{\left|S_{r,e}^{1}(\omega)\right|}$, given that \(S_{r,e}^{3}(\omega)\) typically exhibits the largest magnitude among the higher-order sensitivity functions.

 Although the mentioned analyses can be used to observe the presence of HOSIDFs and compare two different designs, the main question remains: to what extent do open-loop HOSIDFs analysis represent the nonlinearities present in the error (as done in \cite{LukeIFAC2024,karbasizadeh2022band}), and how accurately does the third-order sensitivity capture the overall effect of nonlinearity introduced by the reset element within the closed-loop system (as done in \cite{zhang2024enhancing})?

Thus, in the following section, we present a method for evaluating the effect of the nonlinearity introduced by the reset element on closed-loop performance, accounting for all higher-order harmonics. Our objective is to define a metric with time-domain characteristics that enables us to assess the direct impact of HOSIDFs on the error. Subsequently, we seek to establish a connection between the proposed metric and the frequency-domain HOSIDFs, allowing the metric to be expressed in the frequency domain.

\section{Quantification of Nonlinearity: a metric for robust design of reset control systems.}\label{sec: Robustness factor}

In this section, we introduce the robustness factor in the time domain as a metric to quantify the effect of nonlinearities on the error signal. We then demonstrate that the 2-norm of this metric can be computed entirely in the frequency domain using closed-loop sensitivity functions.
Accordingly, we define the $\sigma_p$ metric as follows:
\begin{definition}
    \label{Definition: sigma p}
    Having $e(t)$ in \eqref{eq ess} as the steady-state error of the closed-loop RCS and $e_1(t)$ as the first-order harmonic of the error ($n=1$ in \eqref{eq en}), we define $\sigma_p(\omega)$ as:
    \begin{equation}
        \label{eq: sigma p}
\sigma_p(\omega)=\frac{\norm{e(t)}_p -\norm{e_1(t)}_p}{\norm{e_1(t)}_p},
    \end{equation}
    where $\norm{f(x)}_{p}$is the $p$-norm ($p\geq 1$) of the $f(x)$ as
    \begin{equation}
        \label{eq: norm p}
        \norm{f(x)}_{p,X}=\left(\int_{X}|f(x)|^p\right)^{1/p},
    \end{equation}
    for complex-valued sequences and functions on $X\subseteq\mathbb{R}^n$.
\end{definition}

Based on Definition \ref{Definition: sigma p}, the contribution of higher-order harmonics to the predicted steady-state error can now be analyzed using any desired norm. Although $\sigma_p(\omega)$ was defined in its most general form, our primary interest lies in gaining deeper insight into the presence and contribution of higher-order harmonics. Therefore, we focus on the 2-norm and employ $\sigma_2$ in this study, as it captures the overall effect of higher-order errors on the steady-state error and provides a more informative indication of variations compared to $e_1(t)$. For example, using the $\infty$-norm only accounts for changes in the signal’s maximum value, which may not effectively reflect magnification effects caused by higher-order harmonics in other parts of the error signal. Consequently, the $\infty$-norm may fail to detect the presence of such errors unless they affect the peak value. Nevertheless, in scenarios where the maximum error is constrained rather than the total error energy, the $\infty$-norm remains useful (see \cite{dastjerdi2022closed} for a more detailed $\infty$-norm analysis of RCSs).

From Definition \ref{Definition: sigma p}, to compute $\sigma_2(\omega)$ and characterize both the magnitude and location of higher-order harmonics effects, it is necessary to reconstruct $e(t)$ and $e_n(t)$ as defined in \eqref{eq ess} and \eqref{eq en}, respectively. However, constructing time-domain signals for each frequency $\omega$ over the interval $t \in (0, T]$, where $T = \frac{2\pi}{\omega}$, is computationally expensive. To address this, we present the following lemma:
\begin{lemma}
    \label{lemma: sigma2}
    Having the 2-norm ($p = 2$) in Definition \ref{Definition: sigma p} yields $\sigma_2(\omega)$, which can be specifically presented in terms of the functions $S_{r,e}^{n}(\omega)$ as follows:
    \begin{equation}
        \label{eq: sigma 2}
        \sigma_2(\omega)= \frac{\sqrt{\sum_{n=1}^\infty |S_{r,e}^{n}(\omega)|^2} - {|S_{r,e}^{1}(\omega)|}}{|S_{r,e}^{1}(\omega)|}.
    \end{equation}
    \textbf{Proof}: See Appendix \ref{App: proof of sigma2}.\\
\end{lemma}

\begin{corollary}
\label{corolary: sigma2 for disturbance}
For the disturbance input case (i.e., from $d_i$ to $e$), the expression for $\sigma_2(\omega)$ given in \eqref{eq: sigma 2} remains valid and accurately quantifies the contribution of higher-order harmonics.\\
\textbf{Proof}: See Appendix \ref{Appendix: proof of corolary}.\\
\end{corollary}

\begin{figure}[!t]
\centering
\includegraphics[width=0.8\columnwidth]{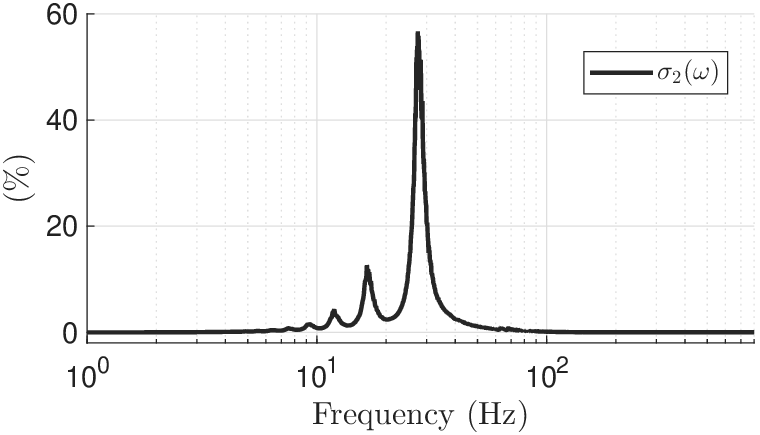}
\caption{The calculated $\sigma_2(\omega)$.}
\label{Fig: sigma2 and ratios}
\end{figure}

Lemma \ref{lemma: sigma2} provides the ratio of the contribution of higher-order errors ($e_3$, $e_5$, $\ldots$, $e_n$) relative to the first-order error ($e_1$). In Fig. \ref{Fig: sigma2 and ratios}, the $\sigma_2(\omega)\%$ is plotted for $C_\text{NL}$. It can be observed that \(\sigma_2(\omega)\) serves as a frequency-domain metric indicating the expected increase in the RMS value of the error resulting from higher-order harmonics at each frequency.

Now, having established a metric that captures the presence and influence of the HOSIDFs on the closed-loop performance, we want to get close to the ideal case. In other words, we aim to have a $\sigma_2(\omega)$ closed to zero, then we should have
\begin{equation*}
    \sqrt{\sum_{n=1}^\infty |S_{r,e}^{n}(\omega)|^2} \approx {|S_{r,e}^{1}(\omega)|},
\end{equation*}
which implies that $|S_{r,e}^{n}(\omega)|_{n>1}\approx 0$. This indicates that higher-order harmonics have no effect on the performance. 

However, it is impossible to eliminate all higher-order harmonics since any reset element exhibiting nonlinearity inherently generates them. Therefore, to take advantage of the reset element while limiting the higher-order harmonics, the next section focuses on developing a method to shape $\sigma_2(\omega)$ at specific frequencies and to a desired extent. This approach allows us to minimize the impact of higher-order harmonics and establish robust performance bounds for the designed reset control system.

\section{Shaping nonlinearity in closed-loop reset control systems}\label{sec: shaping nonlinearity}
In Section \ref{sec: Robustness factor}, we introduced the $\sigma_2(\omega)$ metric, which enables the quantification of the effects of higher-order harmonics on the steady-state error. In this section, our objective is to shape the $\sigma_2$ value to influence the distribution of higher-order harmonics, thereby controlling their impact on the error signal and maintaining a bounded level of nonlinearity induced by the reset element. To achieve this, we first present the pre- and post-filtering approach for the reset element, explaining how it can be employed to shape HOSIDFs. Subsequently, we propose a method for utilizing the $\sigma_2(\omega)$ value in the design of pre- and post-filters.

\subsection{Pre- and post-filtering of the reset element}\label{subsec: pre and post filtering}
In \cite{cai2020optimal}, it is shown that the open-loop HOSIDFs can be altered by choosing different sequences of loop components in the presence of a reset element. Furthermore, in \cite{karbasizadeh2022continuous}, a novel reset control system architecture is proposed, employing a lead filter before, and a lag filter after the reset element to mitigate nonlinearity, forming the "continuous reset element." In this study, we adopt the concept proposed in \cite{karbasizadeh2022continuous}, modeling the reset element as depicted in Fig. \ref{Fig: F and F-1}. This nonlinearity-shaping approach is particularly beneficial, as it does not affect first-order open-loop and closed-loop DFs ($\mathcal{L}_1$ and $S_{r,e}^{1}$). As a result, the fundamental component of the error signal, $e_1(t)$, remains unchanged compared to the unfiltered case. This feature allows us to independently manipulate the higher-order error components ($e_3$, ..., $e_n$), thereby reducing their impact on the overall system performance.

However, in this study, we do not necessarily restrict the pre- and post-filters to lead and lag filters as in \cite{karbasizadeh2022continuous}. Instead, we consider a more general case, where $F$ represents an arbitrary filter placed before the reset element. 
\usetikzlibrary {arrows.meta}
\tikzstyle{sum} = [draw, fill=white, circle, minimum height=0.0em, minimum width=0.0em, anchor=center, inner sep=0pt]
\tikzstyle{block} = [draw,thick, fill=white, rectangle, minimum height=2.5em, minimum width=3em, anchor=center]
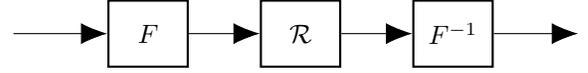
\begin{figure}[!t]
	\centering
	\begin{scaletikzpicturetowidth}{0.85\linewidth}
		\begin{tikzpicture}[scale=\tikzscale]
			\node[coordinate](input) at (0,0) {};
			
			\node[sum] (sum3) at (6.3,0) {};
			

			\node[block] (lead) at (3.2,0) {$\mathcal{R}$};
			\node[block] (controller) at (4.9,0) {$F^{-1}$};
			\node[block] (fo-higs) at (1.5,0) {$F$};
			
			\node[coordinate](output) at (7.2,0) {};

			\draw[arrows = {-Latex[width=8pt, length=10pt]}] (input)  -- node[above]{} (fo-higs);
			\draw[arrows = {-Latex[width=8pt, length=10pt]}] (fo-higs) --node [above]{} (lead);
			\draw[arrows = {-Latex[width=8pt, length=10pt]}] (lead)  --node[above] {} (controller);
			\draw[arrows = {-Latex[width=8pt, length=10pt]}] (controller)  --node[above] {} (sum3);

			
		\end{tikzpicture}
	\end{scaletikzpicturetowidth}
	\caption{The filtered reset architecture.}
	\label{Fig: F and F-1}	
\end{figure}

Therefore, the following lemma provides the modified \(\sigma_2(\omega)\) where a filter \( F \) is implemented before the reset element and its inverse, \( F^{-1} \), is implemented after (as in \eqref{Fig: F and F-1}).

\begin{lemma}
    \label{lemma: filtered sigma}
    Consider the reset element \( \mathcal{R} \) being filtered by \( F \) and \( F^{-1} \) as depicted in Fig.~\ref{Fig: F and F-1}. The value of \( \sigma_2(\omega) \) then changes with respect to the filter \( F \) according to:
    \begin{equation}
    \squeezespaces{0.1}
        \label{eq: filtered sigma}
        \sigma_2^{}(\omega) = \frac{\sqrt{|F(j\omega)|^2 \sum_{\mathrm{} n=1}^\infty \left|F^{-1}(nj\omega)\right|^2 |S_{r,e}^{n}(\omega)|^2} - |S_{r,e}^{1}(\omega)|}{|S_{r,e}^{1}(\omega)|}.
    \end{equation}
\end{lemma}
\textbf{Proof:}\\
By introducing the filters \( F \) and \( F^{-1} \) before and after the reset element, the modified pre- and post-reset controllers become
\[
C_\text{pre}^{'}(j\omega) = C_\text{pre}(j\omega) F(j\omega), \quad 
C_\text{pos}^{'}(j\omega) = C_\text{pos}(j\omega) F^{-1}(j\omega).
\]
Accordingly, the open-loop HOSIDF is given by:
\begin{equation}
    \label{eq: Ln' open loop}
    \mathcal{L}_{n}^{'}(\omega) = G(nj\omega)C^{'}_\text{pos}(nj\omega)H_n(\omega)C^{'}_\text{pre}(j\omega)
    e^{j(n-1)\angle C^{'}_\text{pre}(j\omega)}.
\end{equation}
Substituting the expressions for \( C^{'}_\text{pre}(j\omega) \) and \( C^{'}_\text{pos}(nj\omega) \), we obtain:
\begin{equation}
\begin{split}
    \label{eq: Ln' open loop v2}
    \mathcal{L}_{n}^{'}(\omega) ={} & G(nj\omega)C_\text{pos}(nj\omega)F^{-1}(nj\omega)H_n(\omega) \\
    & \times C_\text{pre}(j\omega)F(j\omega)
    e^{j(n-1)\angle \left(C_\text{pre}(j\omega)F(j\omega)\right)}.
\end{split}
\end{equation}

Thus, for \( \left|\mathcal{L}_{n}^{'}(\omega)\right| \), we have:
\begin{equation}
    \label{eq: |Ln'|}
    \left|\mathcal{L}_{n}^{'}(\omega)\right| = \left|\mathcal{L}_{n}(\omega)\right| \left|F^{-1}(nj\omega)\right| \left|F(j\omega)\right|.
\end{equation}

It is straightforward to show that \( \mathcal{L}_{1}^{'}(\omega) = \mathcal{L}_{1}(\omega) \), and consequently, from \eqref{eq Sn}, it follows that \( S_{r,e}^{'1}(\omega) = S_{r,e}^{1}(\omega) \). Additionally, given that the filtering operation does not affect the base linear system (i.e., \( S_\text{bl}(nj\omega) \) remains unchanged), the expression for \( S_{r,e}^{'n}(\omega) \) becomes:
\begin{equation}
    S_{r,e}^{'n}(\omega) = -\mathcal{L}_n^{'}(\omega) S_\mathrm{bl}(nj\omega) |S_{r,e}^{1}(\omega)| e^{jn \angle{S_{r,e}^{1}(\omega)}}.
\end{equation}
This leads to:
\begin{equation}
    \label{eq: |Sn'|}
    \left|S_{r,e}^{'n}(\omega)\right| = \left|S_{r,e}^{n}(\omega)\right| \left|F^{-1}(nj\omega)\right| \left|F(j\omega)\right|.
\end{equation}
Therefore, substituting the updated first-order (\( S_{r,e}^{'1}(\omega) \)) and higher-order (\( S_{r,e}^{'n}(\omega) \)) closed-loop DFs into the \( \sigma_2(\omega) \) expression in \eqref{eq: sigma 2} yields:
\begin{equation}
\squeezespaces{0.1}
    \label{eq: filtered sigma proof}
    \sigma_2^{}(\omega) = \frac{\sqrt{|F(j\omega)|^2 \sum_{n=1}^\infty \left|F^{-1}(nj\omega)\right|^2 |S_{r,e}^{n}(\omega)|^2} - |S_{r,e}^{1}(\omega)|}{|S_{r,e}^{1}(\omega)|},
\end{equation}
which matches the expression given in \eqref{eq: filtered sigma}.
 \qed \\

As shown in the proof of Lemma~\ref{lemma: filtered sigma}, this filtering method does not alter the first-order open-loop and closed-loop DFs. Consequently, regardless of the specific filter applied to \( F \), the desired first-order DF characteristics of the system remain unchanged. This property enables the independent shaping of the HOSIDFs without affecting the first-order DF.

However, the selection and design of the filter \( F \) to achieve a desired \( \sigma_2(\omega) \) value is not straightforward from \eqref{eq: filtered sigma}. Therefore, in the following part, we present a systematic approach for computing the filter \( F \) based on a given \( \sigma_2(\omega) \), in order to establish a robust nonlinearity bound for RCSs.

\subsection{Shaping $\sigma_2(\omega)$: A Robust design for Reset Control Systems}
In the previous section, we demonstrated the effectiveness of the filtering method in shaping the higher-order harmonics. However, it remains necessary to design the filter \( F \) according to performance requirements. The following theorem provides a guideline for selecting and designing the filter \( F \) such that \(\sigma_2(\omega)\) remains below a specified upper bound \(\sigma_{2,\text{max}}(\omega)\).\\
\begin{theorem}
\label{theorem: Psi N N^-1}
 Let $\sigma_{2,{\mathrm{max}}}(\omega)$ denote an upper bound for $\sigma_2(\omega)$ in \eqref{eq: filtered sigma}. Then, any filter $F$ ensures that $\sigma_2(\omega) \leq \sigma_{2,{\mathrm{max}}}(\omega)$ for all $\omega \in (0, \infty)$, provided that it satisfies the inequality given below:
\begin{equation}
    \label{eq: N upper bound}
    \left|F(\omega)\right|\left|F^{-1}(k_m \omega)\right|\leq\Psi(\omega),
\end{equation}
where $\left|F^{-1}(k_m \omega)\right|=\max\limits_{\mathrm{odd}\,n\in [3,\infty)} \left| F^{-1}(n\omega) \right|$, $k_m\in [3,\infty)$, and
\begin{equation}
    \label{eq: Psi}
\Psi(\omega)=|S_{r,e}^{1}(\omega)|\sqrt{\frac{\sigma^2_{2,{\mathrm{max}}}(\omega)+2\sigma_{2,{\mathrm{max}}}(\omega)}{\sum^{\infty}_{\mathrm{}\, n=3}|S_{r,e}^{n}(\omega)|^2}}.
\end{equation}
\textbf{Proof}: See Appendix \ref{App: proof of Psi}.\\
\end{theorem}

Theorem \ref{theorem: Psi N N^-1} establishes a condition on the filter \( F \), stating that as long as the inequality in \eqref{eq: N upper bound} is satisfied, the value of \( \sigma_2(\omega) \) will not exceed the predefined upper bound \( \sigma_{2,{\mathrm{max}}}(\omega) \).\\

\subsubsection{Example}\label{Example2:}
To validate the method presented in Theorem \ref{theorem: Psi N N^-1}, we consider the reset controller ($C_\text{NL}$) designed in Section \ref{sec: case study}. The resulting $\sigma_2(\omega)$ value for $C_\text{NL}$ is shown in Fig.~\ref{Fig: sigma2 and ratios}, where it goes even above $55\%$ at some frequencies. To ensure a robust reset control design, we impose an upper bound on $\sigma_2(\omega)$, defined as $\sigma_{2,\mathrm{max}}(\omega) = 0.15$. This constraint limits the contribution of higher-order harmonics to the error signal to below $15\%$ of the total error across all frequencies.
Therefore, to determine a suitable filter \( F \) using Theorem~\ref{theorem: Psi N N^-1}, it is necessary to compute \(\Psi(\omega)\) for \(\sigma_{2,\mathrm{max}}(\omega) = 0.15\).

\begin{figure}[!t]
\centering
\includegraphics[width=0.8\columnwidth]{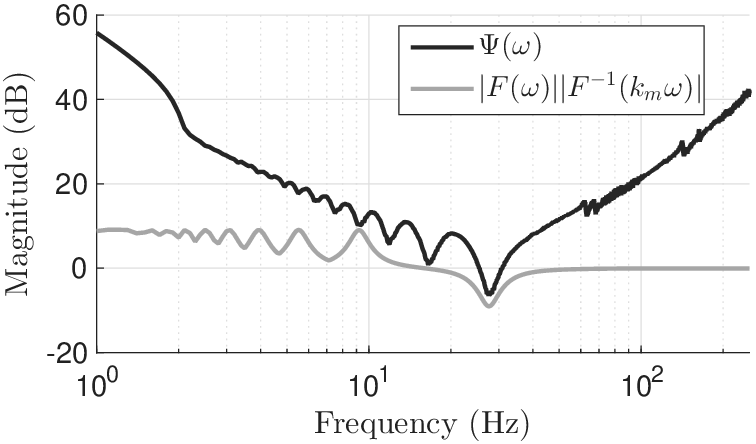}
\caption{The calculated $\Psi(\omega)$ for \( \sigma_{2_{\mathrm{MAX}}}(\omega) = 0.15 \) and the designed filter magnitude characteristic.}
\label{Fig: Psi for 0.01}
\end{figure}
In Fig. \ref{Fig: Psi for 0.01}, the calculated $\Psi(\omega)$ is plotted based on \eqref{eq: Psi}. This function represents an upper bound for $\left|F(\omega)\right||F^{-1}(k_m \omega)|$. In other words, any filter $F$ satisfying the condition $\left|F(\omega)\right||F^{-1}(k_m \omega)| \leq \Psi(\omega)$ will ensure that $\sigma_2(\omega) \leq 0.15$. As shown in Fig. \ref{Fig: Psi for 0.01}, the dip in $\Psi(\omega)$ around the mid-frequency range suggests that a notch filter could be an appropriate choice. It can also be seen from Fig. \ref{Fig: sigma2 and ratios} that $\sigma_2(\omega)$ violates $15\%$ bound only in the mid-frequency range. Then, by selecting $F$ as a notch filter, we can target only those frequencies.

Although any filter \( F \) that satisfies the inequality in \eqref{eq: N upper bound} can, in principle, be selected based on the shape of \(\sigma_2(\omega)\), additional considerations are necessary. Even if a filter \( F \) meets the conditions of Lemma~\ref{theorem: Psi N N^-1}, its placement before the reset element makes it important to avoid introducing excessive gain at high frequencies. This design consideration aligns with the findings of \cite{hosseini2025higherorderharmonics}, which discourage the use of lead-like filters prior to the reset element in order to ensure a smoother and less noisy reset-action signal \( e_r(t) \). Accordingly, we select \( F \) as a notch filter, since it has the potential to satisfy the constraint in \eqref{eq: N upper bound} while also minimizing noise amplification compared to a lead filter.

At this step, we only need to look for a notch filter that will not cross the $\Psi(\omega)$ function. Since $F^{-1}(n\omega)$ will have its peak at $\frac{1}{n^\text{th}}$ of the valley of the $F(\omega)$, we can make sure the $\left|F(\omega)\right||F^{-1}(k_m \omega)|$ value will also stay below $\Psi(\omega)$ everywhere. However, this process still needs a bit of fine-tuning of notch parameters to get a valid $F$ filter. For this example, the filter $F$ is considered as 
\begin{equation}
    \label{eq: notch}
    F(s)= \frac{{s^2}/{\omega_n^2} + {s}/({Q_1 \omega_n}) + 1}{{s^2}/{\omega_n^2} + {s}/({Q_2 \omega_n}) + 1},
\end{equation}
where \( \omega_n, Q_1, Q_2 \in \mathbb{R}_{\geq0} \) with $Q_1=6.79$, $Q_2=2.38$, and $\omega_n=27.5\times 2\pi\,$rad/sec are selected for this example. Having filter $F$ in \eqref{eq: notch}, the value for $\left|F(\omega)\right||F^{-1}(k_m \omega)|$ is visualized in Fig. \ref{Fig: Psi for 0.01} as well. Please note that we calculate the $\left|F(\omega)\right||F^{-1}(k_m \omega)|$ by having $\left|F^{-1}(k_m \omega)\right|=\max\limits_{\mathrm{odd}\,n\in [3,\infty)} \left| F^{-1}(n\omega) \right|$.

With a filter that satisfies the constraint in Theorem \ref{theorem: Psi N N^-1}, we re-implement the controller $C_\text{NL}$ using the same parameters, but now incorporating the pre- and post-filters $F$ and $F^{-1}$. Accordingly, for the closed-loop system depicted in Fig. \ref{Fig: Block diagram CL}, we define the components as follows: $C_\text{pre} = F$, $C_\text{par} = 0$, and $C_\text{pos} = k_\text{c}C_\text{PID}C_\mathfrak{c}F^{-1}$. This results in a filtered version of the controller $C_\text{NL}$.

\begin{figure}[!t]
\centering
\includegraphics[width=0.8\columnwidth]{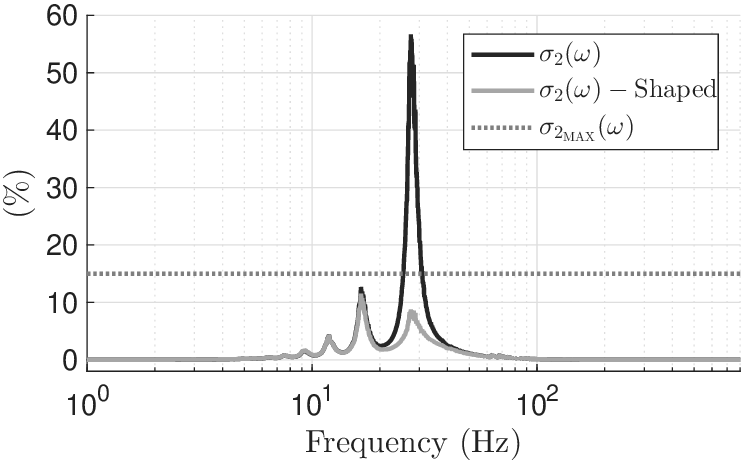}
\caption{The values for $\sigma_{2}(\omega)$ (in percent) for both $C_\mathrm{NL}$ and filtered $C_\mathrm{NL}$ cases.}
\label{fig: sigma 0.01 predicted}
\end{figure}

In Fig. \ref{fig: sigma 0.01 predicted}, the values of $\sigma_2(\omega)$ are presented for both the filtered and non-filtered implementations of the controller $C_\text{NL}$. It can be observed that $\sigma_2(\omega)$ is reduced to below $15\%$ across the entire frequency range, thereby confirming that the designed filter $F$ ensures $\sigma_2(\omega) \leq \sigma_{2,\mathrm{max}}(\omega)$ for all $\omega \in (0, \infty)$.

This improvement has been achieved without modifying any part or parameters of the reset element—simply by applying appropriate pre- and post-filtering. Most importantly, this would not have been possible without the availability of $\sigma_2(\omega)$. The metric $\sigma_2(\omega)$ allows us to bypass the need for individual analysis of HOSIDFs in both open- and closed-loop settings. It provides a frequency-based metric that precisely quantifies the presence and contribution of all higher-order harmonics.

Furthermore, with Theorem \ref{theorem: Psi N N^-1}, there is no longer a need to manually shape HOSIDFs. The theorem offers sufficient insight to directly derive the filter characteristics required to achieve the desired level of robustness against nonlinearities in the system.

\begin{figure}[!t]
\centering
\subfloat[]{\includegraphics[width=0.8\columnwidth]{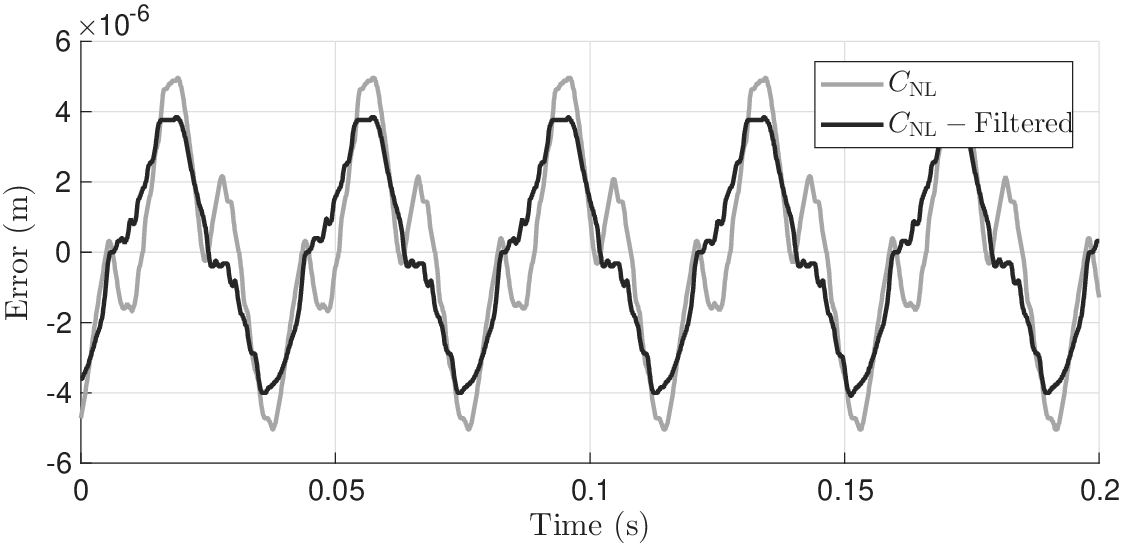}%
\label{fig: Error}}
\hfil
\subfloat[]{\includegraphics[width=0.8\columnwidth]{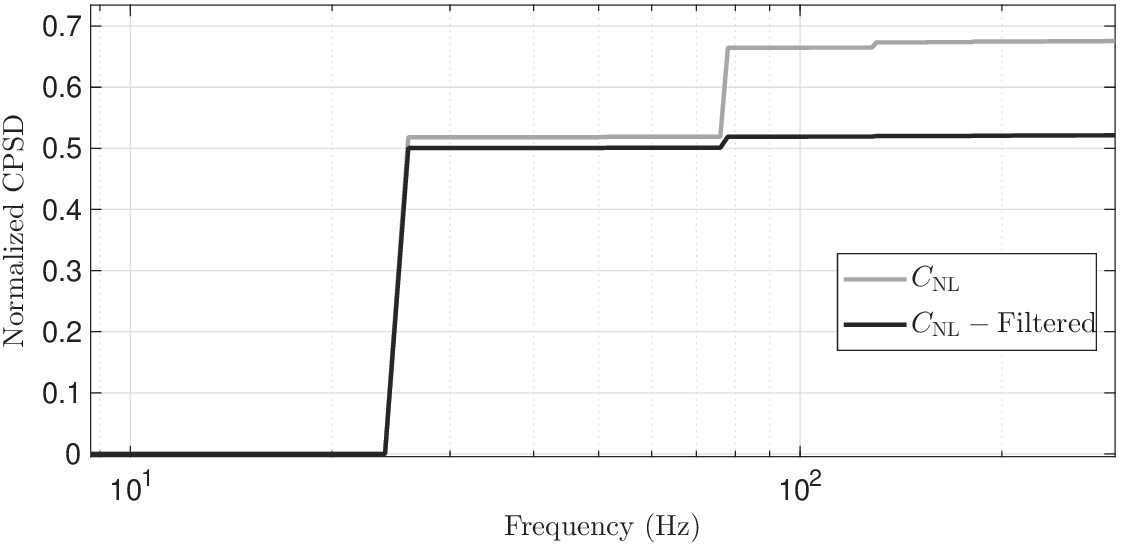}}%
\label{Fig: PSD filtered}
\caption{(a) Measured error signals for $C_\text{NL}$ and filtered $C_\text{NL}$. (b) Normalized CPSD of the measured error signals.}
\label{fig: Error and CPSD}
\end{figure}

Finally, the measured error signals for both $C_\text{NL}$ and the filtered $C_\text{NL}$ are depicted in Fig.~\ref{fig: Error} for the same input as the example in Section \ref{sec: case study}. It can clearly be seen that the unwanted nonlinearities associated with the higher-order harmonics of the reset element have been reduced in the filtered case. To provide a clearer visualization, Fig.~\ref{fig: Error and CPSD}b illustrates the CPSD of the error signals. It is evident that the unwanted jumps in the CPSD, previously attributed to higher-order harmonics in $C_\text{NL}$, are significantly reduced in the filtered case. This indicates a diminished contribution of higher-order harmonics to the error signal, thereby confirming the effectiveness of the proposed nonlinearity shaping approach.
Please note that we analyzed the system in the vicinity of the frequency corresponding to the peak of $\sigma_2(\omega)$ ($28\,$Hz) in order to account for the worst-case contribution of nonlinearity and to explore potential improvements. Moreover, since the implemented filters are notch and inverse-notch filters, their effective operating regions are centered around this frequency, as evidenced by the reduction of $\sigma_2(\omega)$ shown in Fig.~\ref{fig: sigma 0.01 predicted}. Consequently, at both very low and very high frequencies, the two nonlinear systems exhibit similar behavior, whereas around the peak of nonlinearity, the filtered $C_\text{NL}$ demonstrates superior performance in terms of reduced higher-order harmonics.

\begin{remark}
    \label{rem: digital implementation}
    All linear and reset controllers discussed in this study are implemented in a digital framework. The discretization of all LTI elements is performed using the Tustin approximation method, which provides adequate phase preservation of the continuous-time system compared to other approximation techniques, especially up to frequencies close to the Nyquist frequency \cite{aastrom2013computerTustin}. The discrete-time realization of the reset element employs the method proposed in \cite{hosseini2025AddOnFilterDesign}. Further details are provided in \cite[Section V.B]{hosseini2025AddOnFilterDesign}.
\end{remark}

\section{Conclusion}\label{sec: Conclusion}
This paper addressed the challenge of assessing and limiting the nonlinear effects introduced by reset elements in control systems, particularly those stemming from higher-order harmonics. A novel robustness factor, $\sigma_2(\omega)$, was introduced to quantify the contribution of HOSIDFs to the error signal. This metric enables a frequency-domain analysis of reset-induced nonlinearities without requiring time-domain simulations and eliminates the need to individually assess each HOSIDF in both open-loop and closed-loop configurations.

Building on this concept, a systematic filter design method was proposed that allows predefining an upper bound on the acceptable nonlinearity level in the error signal. By designing appropriate pre- and post-filters, the robustness factor can be constrained without altering the DF behavior of the system. However, we showed that it is possible to first define a constraint on the robustness factor and then find the required filter to automatically achieve that level of robustness.

The proposed approach was validated on a planar precision positioning stage, demonstrating how the robustness factor serves as an effective design tool to balance performance gains from reset elements against undesired nonlinear distortions. Experimental and frequency-domain results confirm that the designed filters successfully reduce higher-order harmonic content.

Future work may extend this method to the design of shaping filters, with the objective of identifying a shaping filter that minimizes the robustness factor. Additionally, an optimization problem could be formulated to determine the required filter without relying on a fixed transfer function (such as the notch filter used in this study).

{\appendices
\section{proof of Lemma \ref{lemma: sigma2}}\label{App: proof of sigma2}
Based on Definition \ref{Definition: sigma p} and \eqref{eq: sigma p}, for $\sigma_2(\omega)$ we have:
\begin{equation}
    \label{eq: sigma2 time}
\sigma_2(\omega)=\frac{\norm{e(t)}_2 -\norm{e_1(t)}_2}{\norm{e_1(t)}_2},
\end{equation}
where for a time period of $T$ for $e_1(t)$, from \eqref{eq: norm p} we can write
\begin{equation}
    \label{eq: e1 L2}
\norm{e_1(t)}_{2,T}=\sqrt{\int_{T}{|e_1(t)|^2 \mathrm{d}t}}.
\end{equation}
From \eqref{eq en}, for $n=1$ we have
\begin{equation}
\label{eq: e1 S1}
    e_1(t)=|S_{r,e}^{1}(\omega)|\sin{(\omega t+\angle{S_{r,e}^{1}(\omega)})}.
\end{equation}
Thus, substituting \eqref{eq: e1 S1} in \eqref{eq: e1 L2} for $T=(0,2\pi]$ gives
\begin{equation}
\norm{e_1(t)}_2=\sqrt{\int_{0}^{2\pi}{|S_{r,e}^{1}(\omega)|^2 \sin^2{(\omega t+\angle{S_{r,e}^{1}(\omega)})\mathrm{d}t}}},
\end{equation}
which can be simplified to
\begin{equation}
\label{eq: e1 L2 final}
\norm{e_1(t)}_2=\left|S_{r,e}^{1}(\omega)\right|\sqrt{\pi}.
\end{equation}

For $e(t)$, combining \eqref{eq ess} and \eqref{eq en} gives
\begin{equation}
e(t)=\sum_{n=1}^\infty\left|S_{r,e}^{n}(\omega)\right| \sin(n\omega t+\angle{S_{r,e}^{n}(\omega)}),
\end{equation}
where 2-norm of $e(t)$ can be calculated as
\begin{equation}
\label{eq: e L2 sum}
\norm{e(t)}_2=\sqrt{\int_{0}^{2\pi}\left(\sum_{n=1}^\infty\left|S_{r,e}^{n}(\omega)\right| \sin(n\omega t+\angle{S_{r,e}^{n}(\omega)})\right)^2\mathrm{d}t}.
\end{equation}
By using Parseval’s identity \cite{Parsaval}, for sinusoidal functions $e_n(t)$ we can write
\begin{equation}
\sqrt{\int_{0}^{2\pi}\left(\sum_{n=1}^\infty e_n(t)\right)^2\mathrm{d}t}=\sqrt{\int_{0}^{2\pi}\sum_{n=1}^\infty e^2_n(t)\mathrm{d}t}.
\end{equation}
Thus, for \eqref{eq: e L2 sum} we have
\begin{equation}
\label{eq: e L2 individual}
\norm{e(t)}_2=\sqrt{\int_{0}^{2\pi}\sum_{n=1}^\infty\left|S_{r,e}^{n}(\omega)\right|^2 \sin^2(n\omega t+\angle{S_{r,e}^{n}(\omega)})\mathrm{d}t},
\end{equation}
which is equal to
\begin{equation}
\label{eq: e L2 individual 2}
\norm{e(t)}_2=\sqrt{\sum_{n=1}^\infty\int_{0}^{2\pi}\left|S_{r,e}^{n}(\omega)\right|^2 \sin^2(n\omega t+\angle{S_{r,e}^{n}(\omega)})\mathrm{d}t}.
\end{equation}
Thus for $\norm{e(t)}_2$ we have
\begin{equation}
    \label{eq: e L2 norm final}
\norm{e(t)}_2=\sqrt{\pi\sum_{n=1}^\infty \left|S_{r,e}^{n}(\omega)\right|^2}.
\end{equation}
Finally, substituting \eqref{eq: e1 L2 final} and \eqref{eq: e L2 norm final} in \eqref{eq: sigma2 time} results in
\begin{equation}
        \label{eq: sigma 2 in proof omega}
        \sigma_2(\omega)= \frac{\sqrt{\sum_{n=1}^\infty |S_{r,e}^{n}(\omega)|^2} - {|S_{r,e}^{1}(\omega)|}}{|S_{r,e}^{1}(\omega)|},
    \end{equation}
    which is equal to the one in \eqref{eq: sigma 2}. \qed
\section{Proof of Corollary \ref{corolary: sigma2 for disturbance}}\label{Appendix: proof of corolary}
 In this proof, we show that the $\sigma_2(\omega)$ presented in \eqref{eq: sigma 2} gives the contribution of higher-order harmonics caused by disturbance input ($d_i$) as well. Thus by considering $r(t)=0$, $d_i=\hat{d}\sin{\left(\omega t\right)}$, and $d_n=0$, for error signal $e(t,\omega)$ we can write:
\begin{equation}
    \label{eq ess di}
        e_\text{}(t,\omega)=\sum_{n=1}^{\infty}e_n(t,\omega),
\end{equation}
    where
\begin{equation}
    \label{eq en di}
e_n(t,\omega)=|S_{d,e}^{n}(\omega)|\sin{(n\omega t+\angle{S_{d,e}^{n}(\omega)})},
\end{equation}
with higher-order sinusoidal input sensitivity functions $S_{d,e}^{n}(\omega)$ ($\text{n}^\text{th}$-order HOSIDF calculated from input $d_i$ to the output e) as \cite[Section IV, with $w=d$]{LukeLure}:
\begin{equation}
\label{eq Snd}
S_{d,e}^{n}(\omega) =
\begin{cases}
\displaystyle \frac{G(j\omega)}{1 + \mathcal{L}_1(\omega)}, \hfill \text{for } n = 1, \\[10pt]
\displaystyle -\mathcal{L}_n(\omega) S_\mathrm{bl}(nj\omega) 
\left(|S_{d,e}^{1}(\omega)| e^{jn\angle S_{d,e}^{1}(\omega)}\right), \\
\hfill \text{for odd } n \geq 2, \\[5pt]
0, \hfill \text{for even } n \geq 2.
\end{cases}
\end{equation}
Thus, having $S_{r,e}^{n}(\omega)$ from \eqref{eq Sn} we can write:
\begin{equation}
\label{eq: d proof 1}
    \left|S_{d,e}^{1}(\omega)\right|=|G(j\omega)|\left|S_{r,e}^{1}(\omega)\right|,
\end{equation}
and
\begin{equation}
\label{eq: d proof 2}
    \left|S_{d,e}^{n}(\omega)\right|=|G(j\omega)|\left|S_{r,e}^{n}(\omega)\right|.
\end{equation}
Thus, the $\sigma_2(\omega)$ for disturbance input case can be written as:
\begin{equation}
        \label{eq: sigma 2 in proof d}
        \sigma_2(\omega)= \frac{\sqrt{\sum_{n=1}^\infty |G(j\omega)|^2|S_{r,e}^{n}(\omega)|^2} - {|G(j\omega)||S_{r,e}^{1}(\omega)|}}{|G(j\omega)||S_{r,e}^{1}(\omega)|},
    \end{equation}
which can be simplified as
\begin{equation}
        \label{eq: sigma 2 in proof d2}
        \sigma_2(\omega)= \frac{\sqrt{\sum_{n=1}^\infty |S_{r,e}^{n}(\omega)|^2} - {|S_{r,e}^{1}(\omega)|}}{|S_{r,e}^{1}(\omega)|},
    \end{equation}
    which is equal to the one in \eqref{eq: sigma 2}.
\qed

\section{proof of Theorem \ref{theorem: Psi N N^-1}}\label{App: proof of Psi}
We consider $\sigma_{2,\mathrm{max}}(\omega)$ as the maximum allowed value for $\sigma_{2}(\omega)$ as
\begin{equation}
    \label{eq: sigma max upper}
    \sigma_2(\omega)\leq \sigma_{2,\mathrm{max}}(\omega), \quad \forall \omega \in \mathbb{R}_{>0}.
\end{equation}
Thus, from \eqref{eq: filtered sigma} we should have
    \begin{equation}
    \squeezespaces{1}
    \begin{split}
        \label{eq: max 1}
        \frac{\sqrt{|F(j\omega)|^2 \sum_{n=1}^\infty \left|F^{-1}(nj\omega)\right|^2 |S_{r,e}^{n}(\omega)|^2} - |S_{r,e}^{1}(\omega)|}{|S_{r,e}^{1}(\omega)|}\leq\\
        \sigma_{2,\mathrm{max}}(\omega),
        \end{split}
    \end{equation}
    where we can write
    \begin{equation}
    \begin{split}
|F(j\omega)|^2\sum_{\mathrm{}\,n=1}^\infty \left|F^{-1}(nj\omega)\right|^2|S_{r,e}^{n}(\omega)|^2\leq\\
\left|S_{r,e}^{1}(\omega)\right|^2\left(\sigma_{2,\mathrm{max}}(\omega)+1\right)^2.
\end{split}
    \end{equation}
By extracting $|S_{r,e}^{1}(\omega)|^2$ from $\sum_{\mathrm{}\,n=1}^\infty \left|F^{-1}(nj\omega)\right|^2|S_{r,e}^{n}(\omega)|^2$, it gives
\begin{equation}
\label{eq: ineq app2}
    \begin{split}
|F(j\omega)|^2\sum_{\mathrm{}\,n=3}^\infty \left|F^{-1}(nj\omega)\right|^2|S_{r,e}^{n}(\omega)|^2\leq\\
\left|S_{r,e}^{1}(\omega)\right|^2\left(\sigma_{2,\mathrm{max}}^2(\omega)+2\sigma_{2,\mathrm{max}}(\omega)\right).
\end{split}
    \end{equation}
Please note that even values for $n$ are not considered since from \eqref{eq Sn}, there are no even-order sensitivities for reset control systems.

At this stage we aim to extract $\left|F^{-1}(nj\omega)\right|^2$ from $\sum_{\mathrm{}\,n=3}^\infty \left|F^{-1}(nj\omega)\right|^2|S_{r,e}^{n}(\omega)|^2$ as well. Therefore, we assume at every frequency $\omega$, there is a $n=k_m$ where gives
\begin{equation}
    \label{eq: km}
    \left|F^{-1}(k_mj\omega)\right|=\max\limits_{\mathrm{odd}\,n\in [3,\infty)} \left| F^{-1}(n\omega) \right|,
\end{equation}
for $k_m\in [3,\infty)$. Thus, if \eqref{eq: ineq app2} holds for $\left|F^{-1}(k_mj\omega)\right|^2$ it will hold for all $\left|F^{-1}(nj\omega)\right|^2$ as well. Therefore we can write \eqref{eq: ineq app2} as follows:
\begin{equation}
\label{eq: ineq app2 km}
    \begin{split}
|F(j\omega)|^2\sum_{\mathrm{}\,n=3}^\infty \left|F^{-1}(k_mj\omega)\right|^2|S_{r,e}^{n}(\omega)|^2\leq\\
\left|S_{r,e}^{1}(\omega)\right|^2\left(\sigma_{2,\mathrm{max}}^2(\omega)+2\sigma_{2,\mathrm{max}}(\omega)\right).
\end{split}
    \end{equation}
Now we are able to extract $\left|F^{-1}(k_mj\omega)\right|^2$ from $\sum_{\mathrm{}\,n=3}^\infty \left|F^{-1}(k_mj\omega)\right|^2|S_{r,e}^{n}(\omega)|^2$, which after simplification, results in
\begin{equation}
\begin{split}
         \left|F(j\omega)\right||F^{-1}(k_m j\omega)|\leq\\
|S_{r,e}^{1}(\omega)|^2\sqrt{\frac{\sigma^2_{2,\mathrm{max}}(\omega)+2\sigma_{2,\mathrm{max}}(\omega)}{\sum^{\infty}_{\mathrm{}\, n=3}|S_{r,e}^{n}(\omega)|^2}},
\end{split}
    \end{equation}
where, by having
\begin{equation}
    \label{eq: Psi proof}
\Psi(\omega)=|S_{r,e}^{1}(\omega)|\sqrt{\frac{\sigma^2_{2,{\mathrm{max}}}(\omega)+2\sigma_{2,{\mathrm{max}}}(\omega)}{\sum^{\infty}_{\mathrm{}\, n=3}|S_{r,e}^{n}(\omega)|^2}},
\end{equation}
we have
\begin{equation}
    \label{eq: N upper bound proof}
    \left|F(\omega)\right||F^{-1}(k_m \omega)|\leq\Psi(\omega).
\end{equation}
It means applying any filter $F$ that holds for inequality in \eqref{eq: N upper bound proof}, will result in $\sigma_2(\omega)$ value where 
\begin{equation}
    \label{eq: sigma max upper 2}
    \sigma_2(\omega)\leq \sigma_{2,\mathrm{max}}(\omega), \quad \forall \omega \in \mathbb{R}_{>0}.
\end{equation}
\qed
}
\bibliography{ref}
\bibliographystyle{IEEEtran}
\begin{IEEEbiography}[{\includegraphics[width=1in,height=1.25in,clip,keepaspectratio]{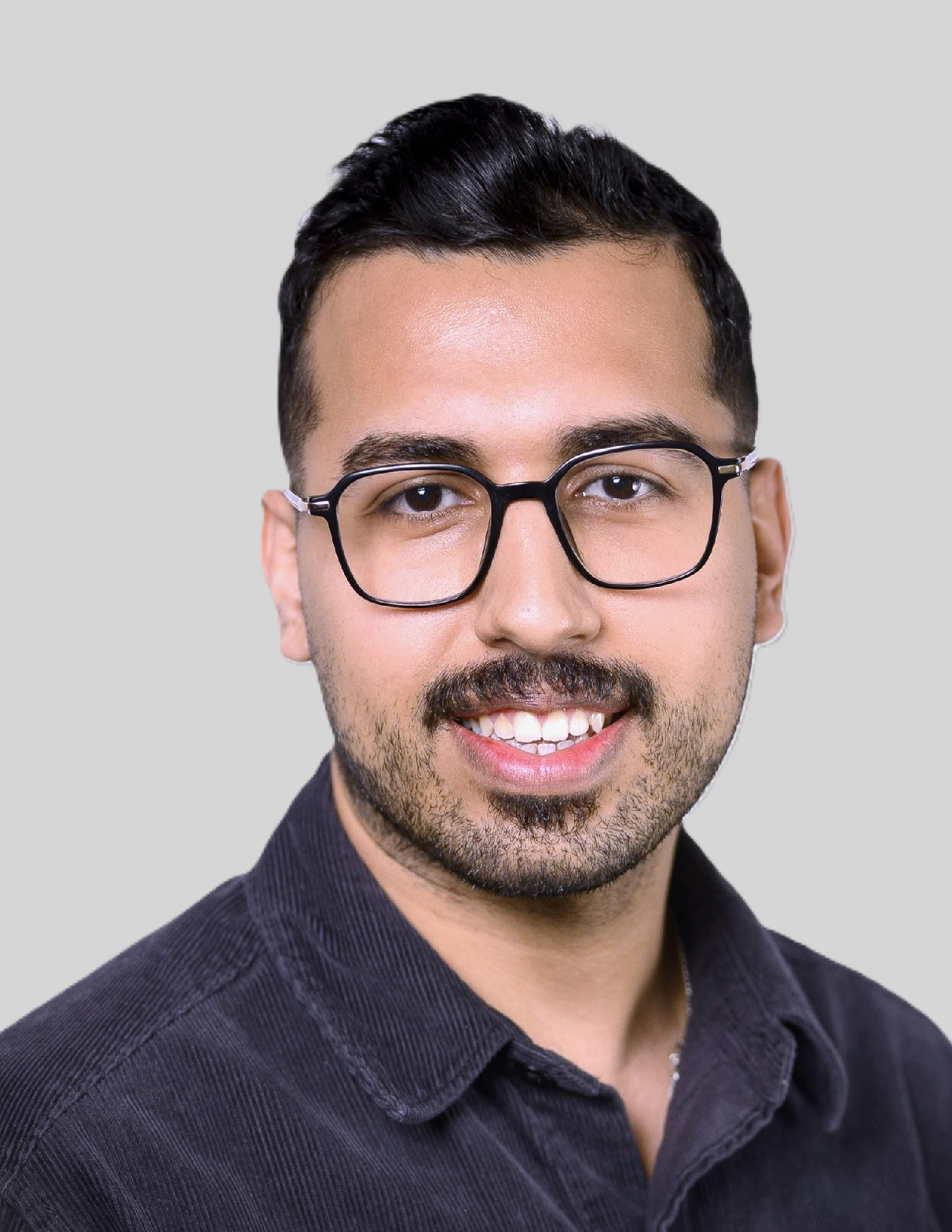}}]{S. Ali Hosseini}
received his M.Sc. degree in Systems and Control Engineering, specializing in nonlinear control (with a focus on hybrid integrator-gain systems), from Sharif University of Technology, Tehran, Iran, in 2022.

He is currently pursuing a Ph.D. in the Department of Precision and Microsystems Engineering at Delft University of Technology, Delft, The Netherlands. His research focuses on addressing industrial control challenges using nonlinear control techniques in close collaboration with ASMPT, Beuningen, The Netherlands. His research interests include precision motion control, nonlinear control systems (such as reset and hybrid systems), and mechatronic system design.

\end{IEEEbiography}

\begin{IEEEbiography}[{\includegraphics[width=1in,height=1.25in,clip,keepaspectratio]{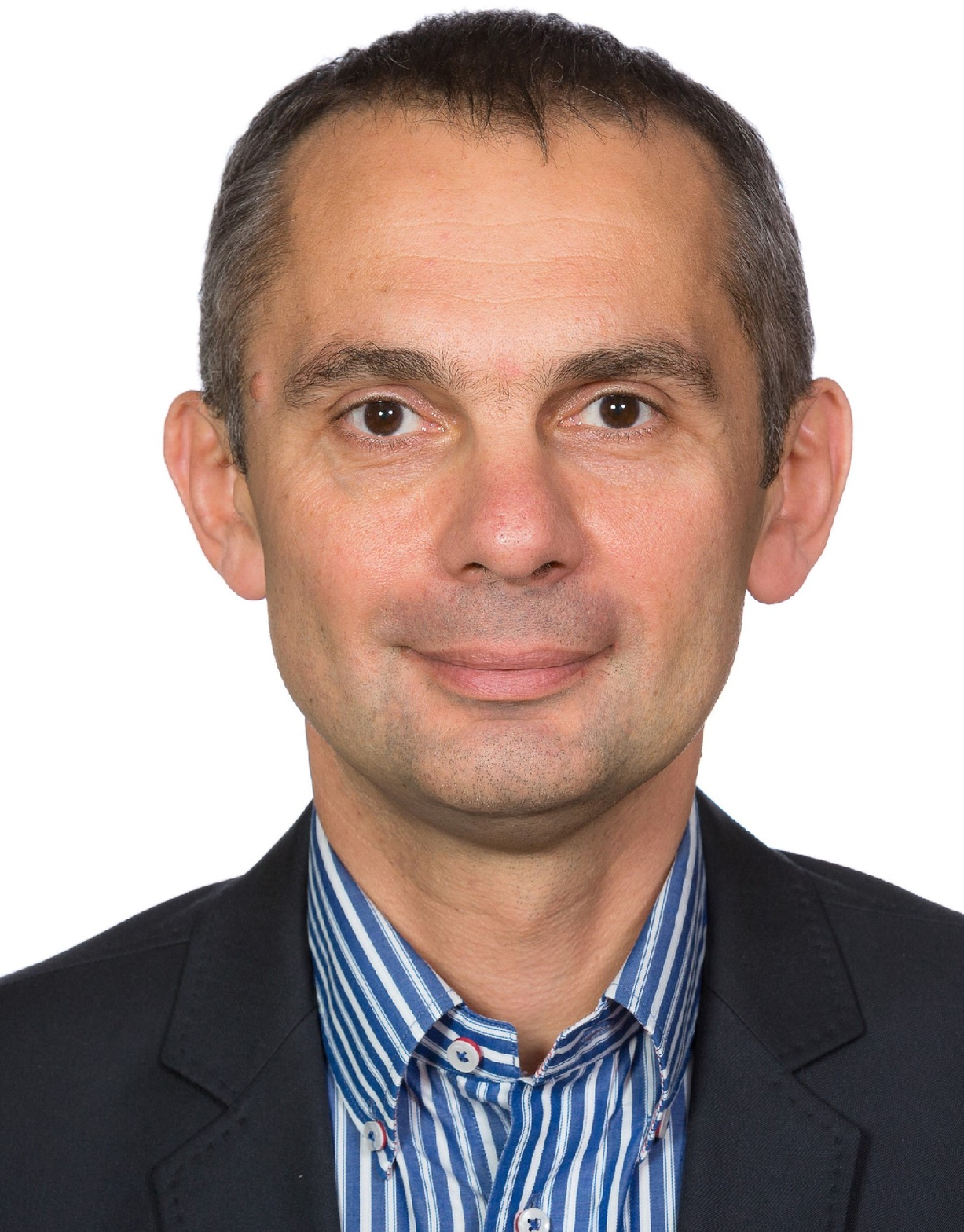}}]{Dragan Kosti\'c}
received the Ph.D. degree in control technology and robotics from the Eindhoven University of Technology, Eindhoven, The Netherlands, in 2004.

His professional positions range from research and teaching at knowledge institutions to professional engineering in commercial companies. Multidisciplinary system modelling and identification, data-based controls, and nonlinear control designs are his main areas of expertise. He works at ASMPT in Beuningen as the Research and Development director for mechatronics. His current research interests include modelling, dynamical analysis, and control of hi-tech mechatronic systems for semiconductor manufacturing.
\end{IEEEbiography}

\begin{IEEEbiography}[{\includegraphics[width=1in,height=1.25in,clip,keepaspectratio]{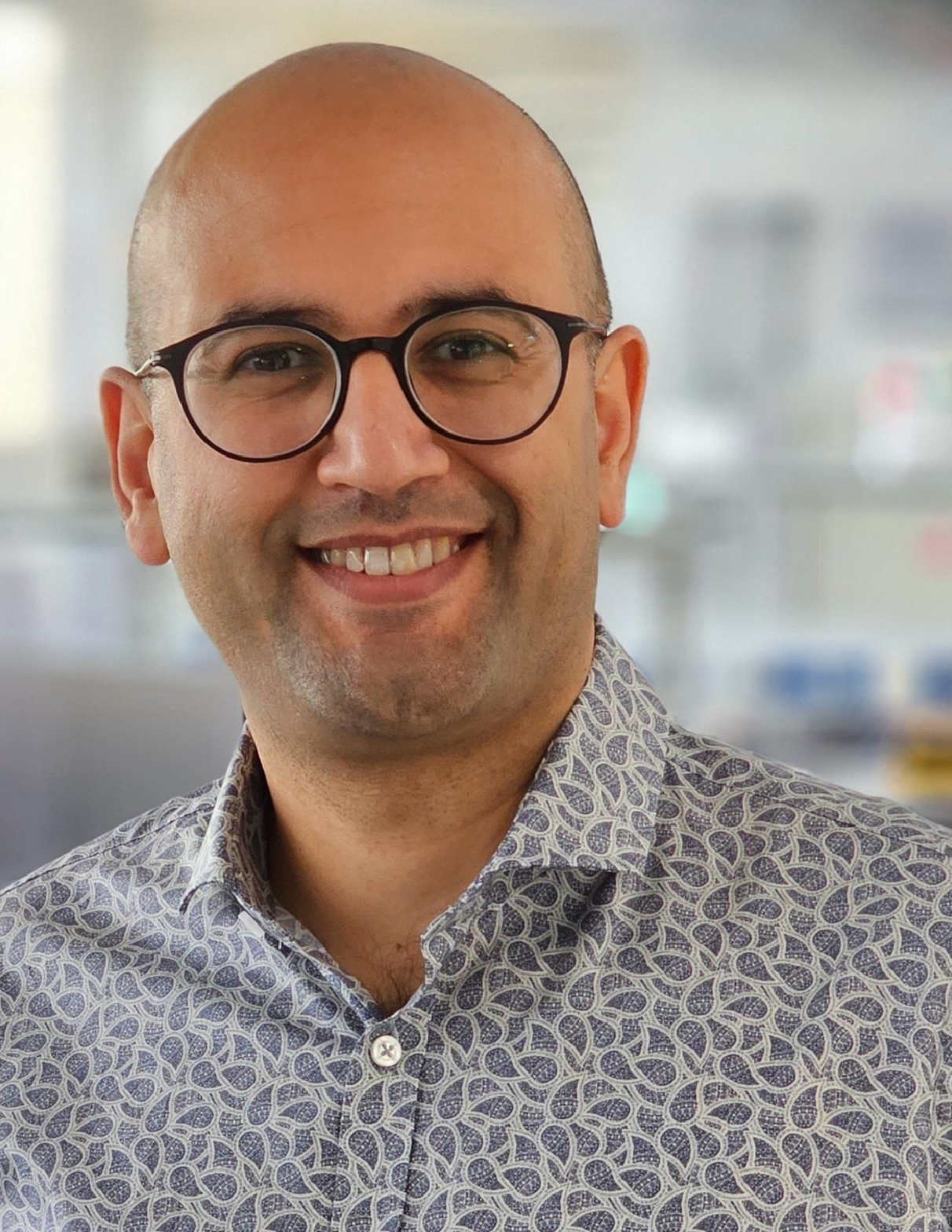}}]{Dr. Hassan Hosseinnia} received his Ph.D. degree cum laude in Electrical Engineering in 2013, specializing in automatic control and its applications in mechatronics. He is currently a faculty member at Delft University of Technology (TU Delft), where he leads advanced research in precision mechatronic system design, high-performance motion control, active vibration damping, and the design and control of electromagnetic and piezoelectric actuators.

His contributions have been recognized internationally. In 2024, he received the Abel Young Scientists Award from the 12th IFAC Conference on Fractional Differentiation and its Applications. His current research focuses on developing innovative control strategies and system architectures to enhance bandwidth and precision in next-generation mechatronic systems. He has published extensively in leading journals and conferences.

Dr. Hosseinnia served as the General Chair of the 7th IEEE International Conference on Control, Mechatronics, and Automation (ICCMA 2019). He also serves on the editorial boards of Mechatronics, Control Engineering Practice, and Fractional Calculus and Applied Analysis. Through his work, Dr. Hosseinnia brings a forward-looking perspective that bridges theoretical depth with industrial applicability, fostering innovation in advanced control and mechatronic system design.
\end{IEEEbiography}



\end{document}